\documentclass[aps,prb,twocolumn,showpacs,superscriptaddress,amsmath,amssymb, amsfonts,10pt]{revtex4-1}
\pdfoutput=1 

\usepackage{dsfont}
\usepackage{graphicx}
\usepackage{color}
\usepackage{xcolor}
\usepackage{enumerate}

\newcommand{\del}{\partial}
\newcommand{\av}[1]{\left\langle #1 \right\rangle}
\newcommand{\ii}{i}
\newcommand{\dd}{\mathrm{d}}
\newcommand{\ee}{\mathrm{e}}

\newcommand{\im}{\text{Im}}
\newcommand{\re}{\text{Re}}
\newcommand{\sgn}{\text{sgn}}
\newcommand{\abs}[1]{\left|#1\right|}
\newcommand{\floor}[1]{\left \lfloor{#1}\right \rfloor }

\newcommand{\LB}{\text{L}}
\newcommand{\TB}{\text{T}}

\newcommand{\imp}{\text{imp}}

\begin{document}

\title{Ultracold quantum wires with localized losses: Many-body quantum Zeno effect}


\author{Heinrich Fr\"{o}ml}
\affiliation{Institute for Theoretical Physics, University of Cologne, 50937 Cologne, Germany}
\author{Christopher Muckel}
\affiliation{Institute for Theoretical Physics, University of Cologne, 50937 Cologne, Germany}
\author{Corinna Kollath}
\affiliation{Physikalisches Institut, University of Bonn, 53115 Bonn, Germany}
\author{Alessio Chiocchetta}
\affiliation{Institute for Theoretical Physics, University of Cologne, 50937 Cologne, Germany}
\author{Sebastian Diehl}
\affiliation{Institute for Theoretical Physics, University of Cologne, 50937 Cologne, Germany}


\begin{abstract}
We study a one-dimensional system of interacting spinless fermions subject to a localized loss, where the interplay of gapless quantum fluctuations and particle interactions leads to an incarnation of the quantum Zeno effect of genuine many-body nature.
This model constitutes a nonequilibrium counterpart of the paradigmatic Kane-Fisher potential barrier problem, and it exhibits strong interaction effects due to the gapless nature of the system. As a central result, we show that the loss probability is strongly renormalized near the Fermi momentum as a realization of the quantum Zeno effect, resulting in a suppression of the emission of particles at the Fermi level. This is reflected in the structure of the particle momentum distribution, exhibiting a peak close to the Fermi momentum.  
We substantiate these findings by three complementary approaches: a real-space renormalization group of a general microscopic continuum model, a dynamical Hartree-Fock numerical analysis of a microscopic model on a lattice, and a renormalization group analysis based on an effective Luttinger liquid description incorporating mode-coupling effects.
\end{abstract}


\pacs{67.85.-d, 64.60.Ht, 71.10.Pm, 64.60.ae} 



\date{\today}
\maketitle


\section{Introduction}

Progress in controllable quantum systems has opened avenues for designing novel states of matter, thanks to the ability to control and engineer both coherent and dissipative processes.  
The interplay of these processes gives rise to phenomena with no counterpart in systems at thermal equilibrium. Examples include driven-dissipative phase transitions~\cite{Carusotto_RMP,Brennecke2013,Tomita2017,Fink2018,Ma2019}, dissipation-induced instabilities~\cite{Dogra2019}, and dissipation-stabilized many-body limit cycles~\cite{Buca2019}.
The quantum Zeno effect (QZE)~\cite{Misra1977} represents perhaps the simplest example of competition between coherent and dissipative processes in quantum systems. In fact, in the context of open quantum systems, measurements can be understood as a coupling to the many degrees of freedom of a bath~\cite{Facchi2002}. 
Recently, experiments with ultracold atoms have directly observed the QZE in many-body systems~\cite{Syassen2008,Barontini2013,Labouvie2016}. In particular, its observation was recently reported for a Bose gas in a quasi one-dimensional optical lattice, subject to a localized dissipation~\cite{Labouvie2016,Muellers2018}, which was accompanied by a number of theoretical works~\cite{Brazhnyi2009,Shchesnovich2010May,Shchesnovich2010Oct, Witthaut2011, Barmettler2011,Zezyulin2012, Kepesidis2012, Kordas2013,Schnell2017,Kiefer2017, Demler2018,Bychek2019,Kunimi2019,Lapp2019,Tonielli2019,Yanay2018,Yanay2020,Krapivsky2019b}.
More recently, experiments with ultracold fermionic wires in the presence of localized dissipation were performed~\cite{Lebrat2019, Corman2019}, studying the interplay between localized losses and transport. Recently, the QZE was also predicted for free fermions subjected to a localized pump~\cite{Krapivsky2019a}.
 
The fundamental mechanism of the QZE in these cases can be understood solely in terms of local, single-body physics: the presence of a fast scale $\gamma$ results in a decoupling of the lossy part from the rest of the system, whose decay is consequently suppressed. 
In Ref.~\onlinecite{Froml2019} the existence of a genuine many-body incarnation of the QZE was demonstrated for a fermionic wire in the presence of a localized loss. The interplay of strong correlations,  gapless quantum modes and a localized impurity was shown to give rise to low-energy collective behaviors, and the system was understood as a dissipative, nonequilibrium analog of the Kane-Fisher problem~\cite{Kane1992,Kane1992Long}. The transport properties across the impurity were shown to be strongly modified: for repulsive interactions, reflection is enhanced, effectively splitting the system into two parts, while for attractive interactions the loss barrier becomes fully transparent. While the fixed points are analogous to the case of the Kane-Fisher problem, a qualitatively different approach of the fixed points was found, affecting the physics near the Fermi momentum $k_F$. These findings were interpreted in terms of a renormalized dissipation strength $\gamma$ in the vicinity of $k_F$. For repulsive interactions, $\gamma$ is infinitely enhanced and losses are suppressed as a consequence of a \emph{fluctuation-induced QZE}. For attractive interactions, instead, $\gamma$ vanishes and the backscattering due to the impurity is suppressed, resulting in a \emph{fluctuation-induced transparency}.

In this work, we further substantiate and extend the results of Ref.~\onlinecite{Froml2019}, by mainly focusing on the identification of observables revealing genuine many-body effects and being sensitive to the QZE. We present results for a number of observables and discuss their robustness against finite system size, finite temperature, and heating effects at the impurity. The analysis of Ref.~\onlinecite{Froml2019} is supplemented with a dynamical Hartree-Fock analysis, and by a second-order renormalization group (RG) analysis for the Luttinger effective description. Our key results are highlighted in the following section, where also the structure of the paper is laid out.

\subsection{Key results}

\emph{Exact solution of the single particle problem ---}
The noninteracting system is studied both numerically and analytically, yielding crucial insights as a basis for the later investigation of the full interacting problem.
The dynamics after a quench of localized loss is found to subdivide into three temporal regimes. In the extensively long second regime a nonequilibrium (quasi)steady state (NESS) is established, providing a well-defined setting for analytical studies. In particular, we study the properties of the NESS in the noninteracting limit by exactly solving the single-particle problem. We obtain exact expressions for density profile, loss rate and currents, revealing the microscopic QZE.
The density profile around the loss site is found to exhibit robust Friedel oscillations, related to the discontinuity in the Fermi distribution and thus a remnant of the initial zero-temperature distribution. Further, Friedel oscillations are known as a key ingredient for severe interactions effects for modes close to the Fermi momentum.
The loss probability $\eta_k$ for particles with momentum $k$ is identified as the key quantity to describe the depletion dynamics of the system. In fact, the wire momentum distribution bears a direct signature of $\eta_k$.
Finally, we derive a Landauer-like formula for the loss rate valid for finite temperatures that allows us to describe the depletion processes by means of the loss probability $\eta_k$, the group velocity $v_k$ and the initial momentum distribution $n_{0,k}$.

\emph{Dynamical Hartree-Fock study ---}
Within a dynamical Hartree-Fock approximation we investigate the impact of interactions on the single particle physics. The Friedel oscillations are found to persist, with a wavelength showing no dependence on the interaction value. The total particle loss rate increases for repulsive interactions and is reduced for attractive ones. The momentum distribution numerically evaluated within this approximation shows a peak close to $k_F$, consistently with the predictions of Ref.~\onlinecite{Froml2019}, which therefore we interpret as a signature of the fluctuation-induced QZE.

\emph{Luttinger liquid and effective temperature ---}
An effective low-energy Luttinger liquid description is used to conduct a perturbative RG analysis to second order in the dissipation strength $\gamma$ within a Keldysh framework. A dual case is identified, for $\gamma \to \infty$, where a perturbative expansion in $\gamma^{-1}$ is viable, thus enlarging the range of applicability of the RG analysis.   
At first order, quantum fluctuations enhance $\gamma$ indefinitely for repulsive interactions, thus suppressing losses and resulting in a fluctuation-induced QZE. For attractive interactions, $\gamma$ vanishes and backscattering is correspondingly suppressed, resulting in a fluctuation-induced transparency. 
At second order, mode-coupling effects are taken into account and an effective temperature is generated at the impurity, cutting off the RG flow. We characterize this self-dephasing regime and identify ranges of values where the first-order effect is visible. 

This article is organized as follows: In Sec.~\ref{sec:model} a microscopic model for the fermionic wire with dissipation is introduced, while in Sec.~\ref{sec:freeSystem} its properties are studied in the absence of interactions. In Sec.~\ref{sec:realSpaceRG} we summarize the real-space RG developed in Ref.~\onlinecite{Froml2019}, while in Sec.~\ref{sec:HF} we present numerical results for the Hartree-Fock dynamics of the wire. In Sec.~\ref{sec:LuttingerLiquid} we develop the Luttinger description for a localized loss, and present the corresponding RG analysis to second order. Finally, in Sec.~\ref{sec:conclusions} the main results are summarized and future directions are discussed.

\section{Model}
\label{sec:model}

We consider spinless fermions with mass $m$ moving in a wire of length $L$ and interacting via a short-range potential $V(x)$, either repulsively or attractively. The system is thus described by the Hamiltonian
\begin{equation}
\label{eq:Hamiltonian-microscopic}
 H =   -\int_x \psi^\dagger(x)\frac{\partial_x^2}{2m}\psi(x) +  \int_{x,y}V(x-y) n(x)n(y),
\end{equation}
with $\psi^\dagger,\psi$ fermionic creation and annihilation operators, $n = \psi^\dagger\psi$ the density operator, and $\int_x = \int_{-L/2}^{L/2} \dd x $. Here and in the rest of the paper we set $\hbar =1$. 

A localized particle loss was implemented in recent experiments with cold atoms either by shining a focused electron beam onto the system~\cite{Barontini2013,Labouvie2016,Muellers2018} or by using a near-resonant optical tweezer~\cite{Lebrat2019,Corman2019}, thus locally exciting atoms resulting in their escape from the trapping potential.
There, this particle loss has been shown to be conveniently described by Markovian single-particle loss: the dynamics of the system can then be described by a quantum master equation~\cite{Zoller_book}
\begin{equation}
\label{eq:master-equation}
\partial_t \rho = - i [H,\rho ] + \int_x \Gamma(x) \left[2 \LB \rho \LB^\dag - \{\LB^\dag \LB, \rho\}\right],
\end{equation}
with $\LB(x) = \psi(x)$ the Lindblad operators and $\Gamma(x) = \gamma \delta(x)$, where $\gamma$ quantifies the dissipation strength. 
We consider the system to be initialized in a thermal state defined by the Hamiltonian~\eqref{eq:Hamiltonian-microscopic} and characterized by a temperature $T$ and a chemical potential $\mu$. At time $t=0$, the localized loss is switched on at position $x=0$, consequently driving the system out of equilibrium. 
\begin{figure}[t]
\centering
\includegraphics[width = 0.4 \textwidth]{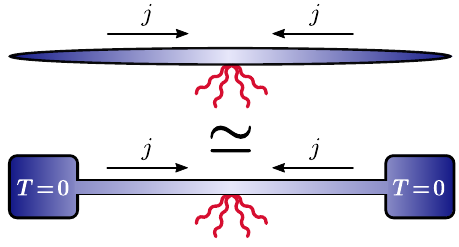}
\caption{%
(Color online). Sketch of a fermionic wire subject to a localized loss. Before emptying out completely, a quasistationary state occurs characterized by constant currents $j$ flowing toward the loss site, analogous to a wire connected to reservoirs at its far ends.
\label{fig:systemsketch} 
}
\end{figure}

In addition to the continuum model~\eqref{eq:Hamiltonian-microscopic}, we also consider a lattice model, defined by the Hamiltonian 
\begin{equation}
\label{eq:Hamiltonian-lattice}
H = -J \sum_j \left(\psi_{j+1}^\dag \psi_j + \text{h.c.} \right) + U \sum_j n_j n_{j+1}, 
\end{equation}
with $\psi_j,\psi^\dagger_j$ the fermionic annihilation and creation operators on site $j$, and $\sum_j = \sum_{j=j_\text{min}}^{j_\text{max}}$. Here, the summation index runs from $j_\text{min} = \floor{-(L-1)/2}$ to $j_\text{max} = \floor{(L-1)/2}$, with $\floor{x}$ the floor function. Periodic boundary conditions are assumed, i.e., $\psi_{j_\text{max}+1} = \psi_{j_\text{min}}$ and ${\psi^\dagger}_{j_\text{max}+1} = {\psi^\dagger}_{j_\text{min}}$. The dynamics is still defined by Eq.~\eqref{eq:master-equation}, where the integral is replaced by a sum over the lattice sites, $\LB(x)$ is replaced by $\LB_j = \psi_j$, and $\Gamma(x)$ is replaced by $\Gamma_j = \gamma \delta_{j,0}$, with $\delta_{i,j}$ the Kronecker delta. 
We set the lattice constant to unity in the rest of the paper.

The reason for considering the lattice Hamiltonian~\eqref{eq:Hamiltonian-lattice} is twofold: on the one hand, it can be regarded as an approximation to the continuum Hamiltonian~\eqref{eq:Hamiltonian-microscopic}, useful for numerical investigations; on the other hand, it represents an interesting model per se, which can be experimentally realized~\cite{Bloch2005, Bloch2008, Gross2017}.

\section{Noninteracting system}
\label{sec:freeSystem} 
We start by considering the effect of the localized loss $\gamma$ on the noninteracting system ($V=0$ and $U=0$ for the continuum and lattice systems, respectively.) On the one hand, this approximation results in a solvable model which allows us to identify the key physical mechanisms and observable quantities of the present problem. On the other hand, it is actually relevant for experiments with ultracold atoms, where the effective particle interaction can be tuned via Feshbach resonances and made negligible~\cite{Lebrat2019, Corman2019}.

\subsection{Temporal regimes and emergence of Friedel oscillations}
\label{sec:timeregimes} 

As a first step, we consider the time evolution of observables in order to gain insight into the remarkably rich dynamics after the quench of the dissipative impurity. Similar nonequilibrium protocols were considered for bosonic systems, leading to a similar dynamics~\cite{Brazhnyi2009,Barmettler2011,Kepesidis2012,Vidanovic2014,Kiefer2017,Bychek2019}. 
This discussion will yield two central building blocks of our analysis: first, the identification of a nonequilibrium (quasi)stationary regime that lasts up to a timescale extensive in the system size, and second, the correlation properties of this nonequilibrium state.

To this end, we consider the dynamics of the lattice model~\eqref{eq:Hamiltonian-lattice} prepared in a state at $T=0$ characterized by a Fermi momentum $k_F = \pi n_0$, with $n_0$ the initial density. All the observables analyzed in this section can be derived from the correlation matrix $C_{ij}(t) = \langle \psi_i^\dagger(t) \psi_j(t) \rangle$, which can be numerically evaluated using the following expression (see Appendix~\ref{sec:appendix_correlations}):
\begin{equation}
\label{eq:correlation}
C_{ij}(t) = \sum_{k,l} G^\dagger_{ik}(t)  G_{jl}^{\vphantom{\dagger}}(t)C_{kl}^{\vphantom{\dagger}}(0),
\end{equation}
where $G(t) = \exp[-i\tilde{H}t]$ is the retarded Green's function, with $\tilde{H}$ the non-Hermitian Hamiltonian associated with the master equation~\eqref{eq:master-equation}, and $C_{ij}(0)$ the correlation matrix evaluated on the initial state. The precise form of $\tilde{H}$ is reported in Appendix~\ref{sec:appendix_correlations}.

We consider the total particle loss rate $\abs{\dd N/ \dd t}$, obtained by evaluating the time derivative of the average total particle number $N(t)$, evaluated in turn from Eq.~\eqref{eq:correlation} as $N(t) = \sum_j C_{jj}(t)$.
In Fig.~\ref{fig:lossrate} the total particle loss rate is shown as a function of time, for different system sizes $L$. Three different regimes then emerge~\cite{Froml2019}:
\begin{figure}[t]
\centering
\includegraphics[width = 0.47 \textwidth]{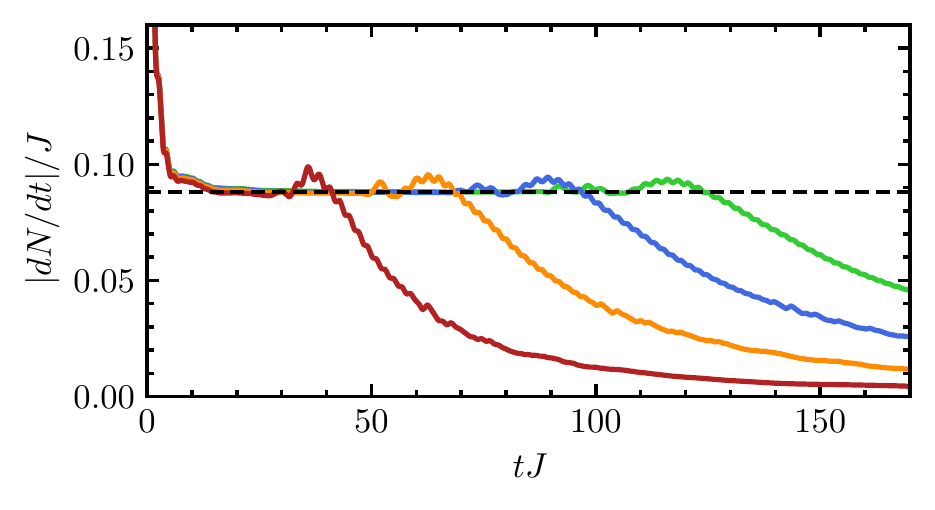}
\caption{%
(Color online). Particle loss rate for the lattice model as a function of time elapsed from the quench, for increasing system sizes, $L = 61, 101, 141, 181$ (leftmost to rightmost curve, respectively). The dashed line indicates the constant loss rate \eqref{eq:lossToN0} obtained analytically for $L \rightarrow \infty$.
For all curves, $\gamma = J, N(0)/L = 0.25$.
\label{fig:lossrate} 
}
\end{figure}
\begin{figure}[t!]
\centering
\includegraphics[width = 0.47 \textwidth]{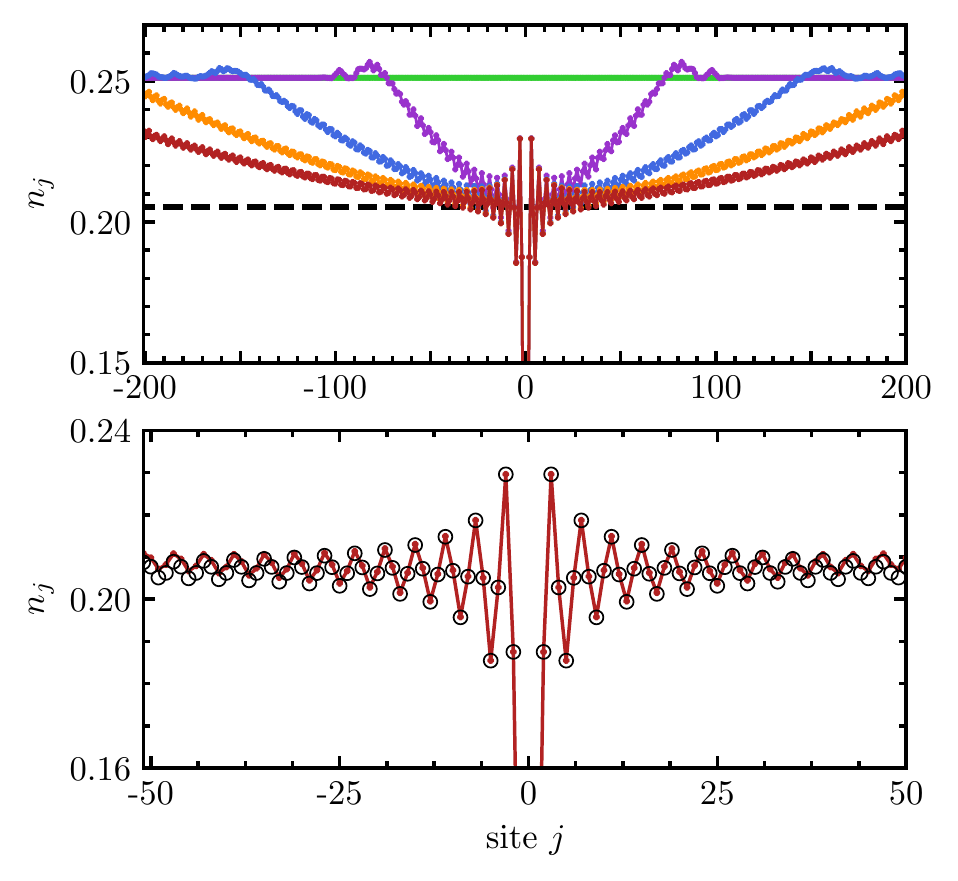}
\caption{%
(Color online). Density profile for the noninteracting lattice model~\eqref{eq:Hamiltonian-lattice}. Upper panel: Density profiles for different times elapsed from the quench ($t J=0,  50, 100, 150, 200$) and analytical average density $n_\text{bg}$ (dashed line).
Lower panel:  Friedel oscillations around the impurity at time $tJ=200$ obtained numerically (solid line) and analytically (hollow circles). 
For all curves $L=601, \gamma = 2 J, N(0)/L = 0.25$.
%
}
\label{fig:friedel} 
\end{figure}

\begin{figure}[t!]
\centering
\includegraphics[width = 0.47 \textwidth]{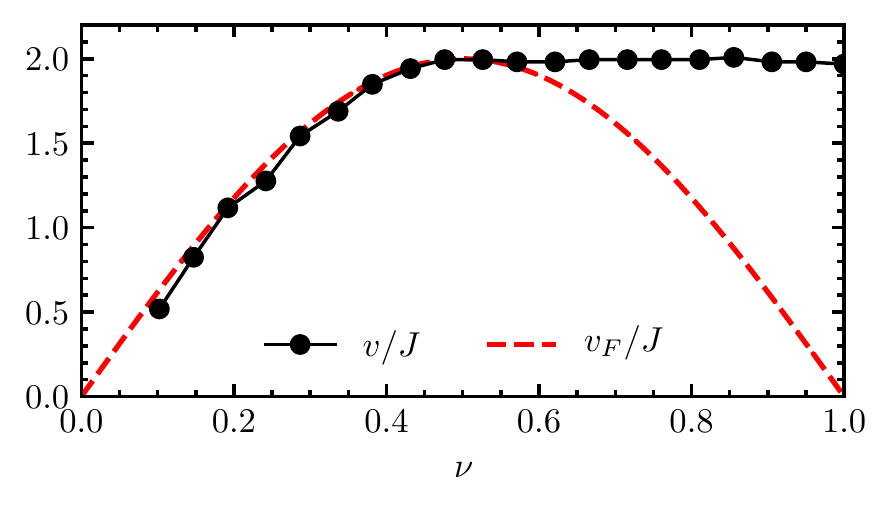}
\caption{%
(Color online).  Speed $v$ of the light cone (dots) separating depleted and unperturbed regions as a function of the initial filling factor $\nu = N(0)/L$, obtained numerically in the lattice system, with $L=401, \gamma = 3 J$. The dashed line indicates the Fermi velocity $v_F$ as a function of the filling factor.
}
\label{fig:lightconeV} 
\end{figure}

\begin{enumerate}[i)]
\item For times $ 0 < t < t_\text{I} \sim \gamma^{-1}$, a rapid drop in the particle number occurs. This drop is caused by a fast, local depletion of the density close to the position of the localized loss.
\item For times $t_\text{I} < t < t_\text{II}$, the particle loss rate is constant. Steady currents (see Sec.~\ref{sec:freelossandcurrents}) are established, directed toward the loss site and feeding on particles of the regions far away from the impurity. The perturbation induced by the loss impurity spreads ballistically with a speed $v$, defining a light cone separating regions affected by the impurity ($ \abs{x} < vt$) from those still unaffected  ($\abs{x} > v t$) (cf. also Fig.~\ref{fig:friedel}, upper panel). 
During this regime, every particle scatters at most once against the impurity without reaching the boundaries, thus rendering the properties of this regime independent of the boundary conditions.  

The speed $v$ hereby corresponds to the one of the fastest particles available in the system. In the continuum model this speed corresponds to the Fermi velocity, $v = v_F$. In the lattice model, due to the nonmonotonic group velocity $v_k = 2 J \sin(k)$, the fastest speed is given by $v = v_F$ below half filling, while for larger fillings it saturates at $v = 2 J$. This is substantiated in Fig.~\ref{fig:lightconeV}, where we show the light-cone velocity as a function of the filling (black dots) in comparison to the corresponding Fermi velocity (red dashed curve). Here, the light-cone velocity is determined numerically by tracking the boundary between depleted and undepleted regions in the density profile, with the depleted region identified by a decrease of the density below $98\%$ of its initial value.

\item  The perturbation eventually reaches the boundaries at a time $t_\text{II} \sim L/(2v)$, and the entire system experiences the presence of the loss. From this point, boundary effects become important, and particles scatter more than once against the loss. These multiple scattering events eventually lead to a full depletion of the system. As the density depletes further, the particle loss rate decreases. We report for completeness that, under certain conditions~\cite{Kepesidis2012}, the wire may possess a dissipationless subspace, namely a set of modes decoupled from the loss preventing the system from completely emptying out. As these conditions are fragile against, e.g., interactions, boundary conditions, or finite size of the impurity, we will not consider this case.  
\end{enumerate}
We emphasize that similar behaviors have been found also in interacting bosonic systems with a localized loss~\cite{Brazhnyi2009,Barmettler2011,Kepesidis2012,Vidanovic2014,Kiefer2017,Bychek2019}, thus suggesting the generality of these dynamical regimes.
The nature of the nonequilibrium steady state realized in the regime $t_\text{I} < t < t_\text{II}$ can be understood as a nonequilibrium (quasi)stationary state (NESS) determined by a balance between the losses at the impurity site and the currents steadily replenishing the region within the light cone (cf.~Fig.~\ref{fig:systemsketch}).

Further insight on this regime can be disclosed by inspecting the dynamics of the density profile. In Fig.~\ref{fig:friedel} (upper panel) the density profile is shown for different times elapsed from the quench: the density at the loss site is strongly depleted, while inside the light cone ($\abs{x} \lesssim v t$) the density approaches a stationary profile, oscillating around a mean value different from the initial value. An analytical expression for this profile is provided in Sec.~\ref{sec:FreeDensityProfile}. In Fig.~\ref{fig:friedel} (lower panel) the comparison between the density profile at the latest time in Fig.~\ref{fig:friedel} (upper panel) to the analytical density profile \eqref{eq:densityProfile} is shown: the density relaxes to its stationary value proceeding from the sites near the loss.
The density oscillations are characterized by a period of $\pi/k_F$, with $k_F$ the Fermi momentum of the initial state, and a further numerical inspection of the asymptotic density profile (not shown), supported by an analytical solution, see Sec.~\ref{sec:FreeDensityProfile}, shows that they decay in space as $\sim 1/x$: they can thus be regarded as Friedel oscillations \cite{Egger1995}. Since these oscillations are characteristic of Fermi systems at $T=0$, their presence and robustness in a nonequilibrium steady state is remarkable, and can be traced back to the dynamics of the particle momentum distribution, which is analyzed in detail in Sec.~\ref{sec:nK}.

Given the extensively large timescale $t_\text{II}$, and the emergence of Friedel oscillation therein, in the following we will focus on the properties of this quasistationary steady state. 

\subsection{Analytical results via Green's function method}
\label{sec:GreensMethod}

Having identified the temporal regime of interest, we substantiate and extend the results discussed in the previous section with exact analytical results. In the present section we will provide the main steps of the derivation, while the discussion of observables and interpretation of the final results will be given in subsequent sections. 
For the sake of brevity, calculations are presented for the continuum system described by the Hamiltonian~\eqref{eq:Hamiltonian-microscopic}. The adaptation to the lattice case is straightforward and presented in Appendix~\ref{sec:S_lattice}.

A major simplification arises when the thermodynamic limit $ L \to \infty $ is considered. However, this limit has to be taken carefully: it does not commute with the limit $t \to \infty$, as a consequence of the fact that the timescale $t_\text{II}$, characterizing the end of the second regime, scales with $L$, as discussed in the previous section. 
In fact, if the limit $L\to\infty$ is taken \emph{after} $t\to\infty$, the system will be in the trivial vacuum state. In contrast, if the limit $L\to\infty$ is taken \emph{before} $t\to\infty$, the system will never reach the end of the second regime, which becomes thus a true stationary state. We will therefore focus on the latter case.

Since in the present section we neglect interactions, the fluctuations of the system are Gaussian and, therefore, the knowledge of the two-times correlation function $C(x,x',t,t') = \langle \psi^\dagger(x,t)\psi(x',t') \rangle$ is sufficient to determine any observable. Exactly as in Eq.~\eqref{eq:correlation}, $C(x,x',t,t')$ can be determined from
\begin{equation}
\label{eq:correlation-retarded}
C(x,x',t,t') \!=\! \int_{y,y'} \!\!\!\! G^*(x,y,t)G(x'\!,y'\!,t') C(y - y'\!,0,0)	, 
\end{equation}
where $G(x,y,t) = -i\theta(t)\langle  \{\psi(x,t),\psi^\dagger(y,0)\}\rangle $ is the retarded Green's function and  
$C(x-x',0,0)$ is the correlation function of the initial state, given by $C(x,0,0) = \int_k \ee^{i k x} n_{0,k}$, with $n_{0,k}$ the initial momentum distribution and $\int_k = \int_{-\infty}^{+\infty} \dd k/(2\pi)$. In order to compute $C(x,x',t,t')$ it is then necessary to evaluate the retarded Green's function. Since $G(x,y,t)$ only depends on the time difference, it is convenient to consider its Fourier transform $G(x,y,\omega)$ in frequency $\omega$. Its explicit form can be derived by solving a Dyson equation (see Appendix~\ref{app:GreensFunction}) and reads
\begin{equation}
 G(x,y,\omega)  = \frac{1}{i} \sqrt{\frac{m}{2\omega}}\left[ \ee^{i\sqrt{2 m \omega}\abs{x-y}} + r(\omega)\,  \ee^{i\sqrt{2 m \omega}(\abs{x}+\abs{y})} \right],
\label{eq:Gxx}
\end{equation}
with
\begin{equation}
r(\omega) = \frac{-\gamma}{\sqrt{2 \omega / m} + \gamma}.
\label{eq:romega} 
\end{equation}
Having provided an explicit form of the retarded Green's function, one can then evaluate Eq.~\eqref{eq:correlation-retarded} analytically in the stationary regime, by taking the limit $t,t' \to +\infty$ while $t-t'$ remains finite, where one obtains (see Appendix~\ref{sec:correlationFunctions}):
\begin{equation}
C(x,x'\!,t,t') \!=\! \int_k\! f^*(\epsilon_k,k, x) f(\epsilon_k,k, x') \ee^{\ii \epsilon_k (t-t')} n_{0,k}	,
\label{eq:correlationsfree} 
\end{equation}	
with $f(\omega,k, x) = \ee^{-ikx} + r(\omega)\ee^{i\sqrt{2 m \omega}|x|}$ and $\epsilon_k= k^2/2m$ the dispersion relation. Notice that the right-hand side of Eq.~\eqref{eq:correlationsfree} only depends on the difference $t-t'$ as a consequence of stationarity in time.  
From the previous equation, a number of exact results can be obtained, which will be discussed in the next sections.

\subsection{Loss probability}
\label{sec:loss_probability}

Beyond providing the technical foundation of the further analysis, the derivation illuminates the physical mechanisms underlying the results.
The retarded Green's function in Eq.~\eqref{eq:Gxx} can be interpreted in the following way. If a particle with momentum $k>0$ and energy $\omega = \epsilon_k$ is injected at $y \ll x$, $G(x,y,\omega)$ will describe the probability amplitude of finding the particle at a position $x$:
\begin{equation}
G(x,y \ll x,\omega) \propto 
\begin{cases}
\ee^{ikx} + r(\omega)\ee^{-ikx} & x<0 \\
t(\omega)\ee^{ikx} & x > 0	,
\end{cases} 	
\label{eq:GScattering} 
\end{equation}
with $t(\omega)=1+r(\omega)$ fixed by the continuity condition at $x=0$. Equivalently, we can understand Eq.~\eqref{eq:GScattering} as the response to a plane wave perturbation incoming from the left with momentum $k$ and energy $\omega = \epsilon_k$ (see Appendix~\ref{app:GreensFunction}).
In other words, $G(x,y \ll x,\omega)$ describes a one-dimensional scattering problem. From this, one can read off the transmission and the reflection amplitudes,  $t_k\equiv t(\omega = \epsilon_k)$ and $r_k \equiv r(\omega = \epsilon_k)$, respectively:
\begin{equation}
\label{eq:scatteringCoeff} 
t_k = \frac{\abs{v_k}}{\abs{v_k} + \gamma}, \quad  r_k = -\frac{\gamma}{\abs{v_k}+ \gamma} ,
\end{equation}
with $v_k= \partial \epsilon_k/\partial k = k/m$ the group velocity.
In contrast to the case of a delta potential barrier, these amplitudes are real-valued and violate probability conservation as $|r_k|^2+|t_k|^2 \neq 1$: a lossy impurity acts as a nonunitary scatterer. The study of this scattering problem yields crucial insights to characterize the losses in the system.

\begin{figure}[t]
\centering
\includegraphics[width = 0.47 \textwidth]{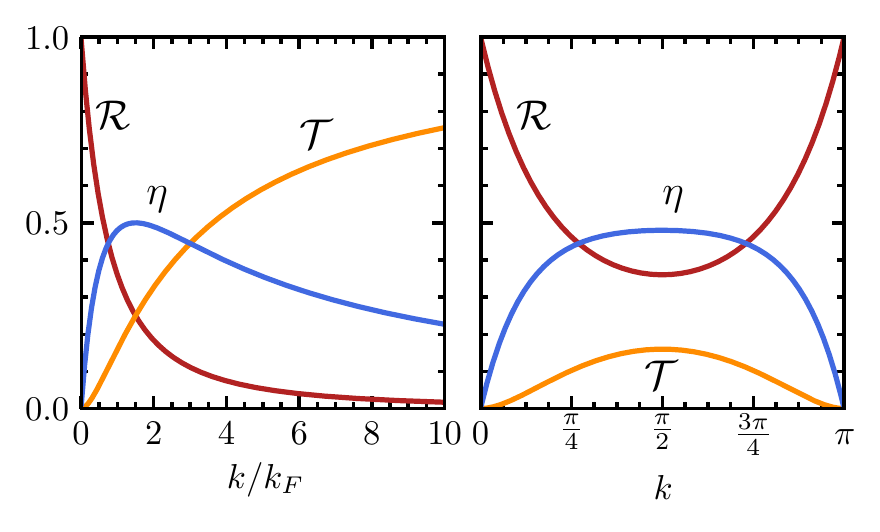}
\caption{%
(Color online). Transmission probability $\mathcal{T}_k$, reflection probability $\mathcal{R}_k$, and loss probability $\eta_k$ as a function of momentum $k$ for the continuum (left panel, $\gamma/v_F = 1.5$) and lattice model (right panel, $\gamma = 3 J$). 
}
\label{fig:scatteringProb} 
\end{figure}

In order to quantify the loss of unitarity, it is then convenient to introduce the \emph{loss probability} $\eta_k$ as
\begin{equation}
\label{eq:etadef} 
\eta_k = 1 - |r_k|^2-|t_k|^2,
\end{equation} 
corresponding to the probability of a particle with momentum $k$ and energy $\epsilon_k$ to be absorbed by the localized loss. By using the definition~\eqref{eq:etadef} and Eq.~\eqref{eq:scatteringCoeff}, the loss probability of the noninteracting system reads
\begin{equation}
\label{eq:defeta} 
\eta_k = \frac{2\gamma \abs{v_k}}{(\gamma+|v_k|)^2}		.
\end{equation}
From Eq.~\eqref{eq:defeta} it can be thus inferred that $\eta_k$ is a nonmonotonic function of the ratio $\gamma/v_k$, corresponding to a manifestation of the QZE. We emphasize that this is not a many-body QZE, but rather a single-particle property.
In Fig.~\ref{fig:scatteringProb} the transmission and reflection probabilities, given by, respectively,
\begin{equation}
\label{eq:scatteringProb} 
\mathcal{T}_k = |t_k|^2, \qquad \mathcal{R}_k = |r_k|^2,
\end{equation}
and the loss probability $\eta_k$ are depicted as a function of momentum $k$ for both the continuum and the lattice model (see Appendix~\ref{sec:S_lattice} for the latter).
The loss probability is nonmonotonic as a function of the momentum $k$ as a consequence of the QZE: more precisely, $\eta_k$ becomes negligible for $|v_k| \ll \gamma$, indicating that particles are rather reflected back than lost from the system. 

The loss probability $\eta_k$ will play a central role in our analysis: it appears ubiquitously in the observables discussed in the following sections and represents a directly observable quantity itself (see Sec.~\ref{sec:nK}). Moreover, $\eta_k$ can be strongly affected by interactions, resulting in a universal behavior at low momenta (see Sec.~\ref{sec:realSpaceRG}).

As a final remark, we notice that the scattering solutions~\eqref{eq:GScattering} can be equivalently obtained by solving a single-particle Schr{\"o}dinger equation with a non-Hermitian Hamiltonian. This equivalence carries over to the lattice system, see Appendix~\ref{sec:S_lattice}. On the lattice, the eigenfunctions of the non-Hermitian Hamiltonian include, besides scattering solutions, a localized state, whose presence is, however,  irrelevant to the effects discussed in this work. The nature of this localized state is discussed in Appendix~\ref{sec:S_lattice_bound_state}. Remarkably, the emergence of the localized state is related to an exceptional point~\cite{Heiss2012} of the non-Hermitian Hamiltonian.

\subsection{Density profile}
\label{sec:FreeDensityProfile} 
\begin{figure}[t]
\centering
\includegraphics[width = 0.47 \textwidth
]{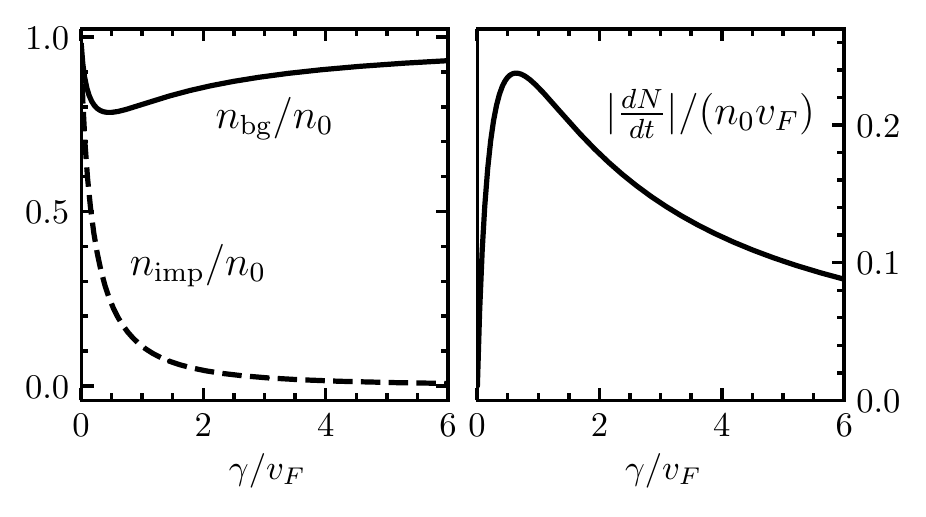}
\caption{%
Left panel: Background density $n_\text{bg}$ (solid) and density at the loss site $n_\text{imp}$ (dashed) in the second temporal regime as a function of the dissipation strength $\gamma$ in the continuum model.
Right panel:
Particle loss rate $|dN/dt|$ in the second regime as a function of the dissipation strength $\gamma$ in the continuum model.
}
\label{fig:nNessnImp} 
\end{figure}

The density profile in the stationary state is defined as $n(x) = C(x ,x, t,t)$ (notice that the time dependence drops out because of time-translational invariance), and, with use of Eq.~\eqref{eq:correlationsfree}, it reads
\begin{align}
n(x) &  = \int_k\left[ 1+r_k^2 + r_k + r_k \cos(2k|x|) \right]n_{0,k} \nonumber\\
& \equiv n_\text{bg} + \delta n(x)		.
\label{eq:densityProfile} 
\end{align}
The density profile thus consists of a homogeneous background density $n_\text{bg}$, corresponding to the mean density in the stationary state, with density modulations $\delta n(x)$. Assuming the system to be prepared at $T=0$, so that $n_{0,k} = \theta(k_F^2 -k^2)$, the behavior of these modulations for $|x|\gg k_F^{-1}$  is given by
\begin{equation}
\label{eq:FriedelOsc} 
\delta n(x) \simeq r_{k_F} \frac{\sin(2k_F|x|)}{2\pi|x|}.
\end{equation}
Hence, the modulations are identified as Friedel oscillations of the fermionic density, characterized by a period $\pi / k_F$ and spatial decay of $\sim 1/x$, confirming the numerical results discussed in Sec.~\ref{sec:timeregimes} (see Fig.~\ref{fig:friedel}). The background density $n_\text{bg}$ can be expressed in terms of $\eta_k$ and evaluated in a closed form:
\begin{align}
\label{eq:nNESSeta} 
n_\text{bg} &= \int_k \left( 1- \frac{\eta_k}{2} \right) n_{0,k} 	.
\end{align}
In addition, the  expression for the density at the impurity site $n_{\imp} = n(x=0)$ can be derived from Eq.~\eqref{eq:densityProfile} and reads
\begin{align}
\label{eq:nimp} 
n_\text{imp} & = \frac{1}{2\gamma} \int_k \, |v_k| \, \eta_k n_{0,k}.
\end{align}
The values of $n_\text{bg}$ and $n_\text{imp}$ for $T=0$ can be found analytically and are shown in Fig.~\ref{fig:nNessnImp} (left panel) as functions of $\gamma/v_F$. The background density $n_\text{bg}$ exhibits a nonmonotonic behavior representing a manifestation of the QZE. On the other hand, the density $n_\text{imp}$ exhibits a monotonically decreasing behavior, as this site remains directly affected by the dissipation for any value of $\gamma$.

\subsection{Particle loss rate and currents}
\label{sec:freelossandcurrents} 

The total particle loss rate $\abs{\dd N/ \dd t}$ is determined by the density at the loss site $n_{\imp}$ via the equation
\begin{equation}
\frac{\dd N}{\dd t} = - 2 \gamma n_{\imp},
\label{eq:lossToN0} 
\end{equation} 
which can be readily obtained by calculating the equation of motion for the average number of particles $N=\langle \hat{N}\rangle$ from Eq.~\eqref{eq:master-equation}. Its analytical expression in the nonequilibrium steady state can be evaluated by using Eq.~\eqref{eq:nimp}. The particle loss rate as a function of $\gamma$ is depicted in Fig.~\ref{fig:nNessnImp} (right panel) and exhibits a nonmonotonic behavior, reflecting once again the QZE: for small dissipation strength $\gamma$ compared to $v_F$, the loss rate increases as $ \sim \gamma$, while at large $\gamma$ it is found to decrease as $\sim \gamma^{-1}$.

The stationary regime is further characterized by steady currents which originate at the far ends of the wire and flow toward the loss site (cf. Fig.~\ref{fig:systemsketch}). These steady currents are calculated as
\begin{equation}
j(x) \!=\! \frac{1}{m} \im  \av{\psi^\dagger(x) \partial_x \psi(x)} \!=\! \frac{1}{m} \im \partial_y C(x,y)\biggr|_{y=x}	 \! 	,
\end{equation}
where the two-point correlations $C(x,y) \equiv C(x,y,t,t)$ in the stationary regime can be obtained from  Eq.~\eqref{eq:correlationsfree}. A straightforward calculation yields (see Appendix~\ref{sec:S_currents})
\begin{equation}
j(x)  = - \sgn(x)  \gamma\, n_\text{imp},
\label{eq:current} 
\end{equation}
with $n_\text{imp}$ given in Eq.~\eqref{eq:nimp}. The sign function entails that the currents are flowing toward the loss site, as expected (we chose, as convention, $j(x)>0$ corresponding to a current flowing to the right). Notice that Eq.~\eqref{eq:lossToN0} can also be straightforwardly derived by integrating the continuity equation
\begin{equation}
\frac{\partial n(x,t)}{\partial t} = \partial_x j(x,t) - 2\gamma\, n(x,t) \delta(x)
\end{equation}
in the stationary state.

%
\subsection{Momentum distribution dynamics}
\label{sec:nK} 
In the presence of a localized loss, particles which scatter against the impurity can either remain in the wire or get lost. Consequently, one expects the momentum distribution of particles in the wire to bear a characteristic signature of the losses. In this section, we analyze the dynamics of the momentum distribution, and show that it is closely related to the loss probability $\eta_k$ defined in Eq.~\eqref{eq:etadef}. Since the momentum distribution is accessible in time-of-flight measurements with cold atoms, it provides a direct way of measuring $\eta_k$. 

The momentum distribution $n_k(t)$ can be evaluated from
\begin{equation}
\label{eq:momentum_distro_def}
n_k(t)  = \frac{2\pi}{L} \langle   \psi^\dagger(k,t) \psi(k,t) \rangle, 
\end{equation}
with $\psi^\dagger(k)$ and $\psi(k)$ the fermionic creation and annihilation operators for a particle with momentum $k$ and $L$ the volume of the system.
In the following, we will focus on two different momentum distributions: the distribution $n_k(t)$ for a finite size $L$ of the system and evaluated at a finite time $t$ elapsed from the quench, and the one for the NESS discussed in Sec.~\ref{sec:freeSystem}, which we dub $n_\text{ness}(k)$.
In particular, $n_k(t)$ describes the momentum distribution of the \emph{entire} system, including the regions outside the light cone (see Sec.~\ref{sec:timeregimes} and Fig.~\ref{fig:friedel}). In contrast, $n_\text{ness}(k)$ describes the momentum distribution only in the region inside the light cone.
These two different cases do not simply coincide in the $L,t \to \infty$ limit as a consequence of the limiting procedure (cf. discussion in Sec.~\ref{sec:GreensMethod}), but they effectively correspond to two different physical quantities. We will elaborate more on this difference later on.

\subsubsection{Momentum distribution for finite $L$ and $t$}

We first consider the momentum distribution $n_k(t)$ of a finite-sized system within the second temporal regime.
A straightforward computation (see Appendix~\ref{sec:S_momentumDistribution}) leads to 
\begin{equation}
n_k(t ) =  n_{0,k} \left[ 
1 -  \frac{|v_k|  t}{L} \eta_k
\right] 
\label{eq:nkt}	,
\end{equation}
which is valid up to order $O(L^{-1})$ and for $t < t_\text{II}$ (see Sec.~\ref{sec:timeregimes}). The linear depletion in time of the modes described by Eq.~\eqref{eq:nkt} has a transparent physical interpretation. The density of particles lost, $n_{0,k} - n_k(t)$, is in fact proportional to three factors, i.e., the original number of particles $n_{0,k}$, the fraction of particles which have reached the impurity at a given time $|v_k| t /L$, and the loss probability $\eta_k$. As a consistency check, we notice that Eq.~\eqref{eq:nkt} renders, after integrating over all momenta $k$, the expected result for the total number of particles, i.e., $N(t) = N(0) +  (\dd N/\dd t) t$, with 
\begin{equation}
 \frac{\dd N}{\dd t} = - \int_k
  \abs{v_k}  \eta_k n_{0,k},
\label{eq:NdotT} 	
\end{equation}
in agreement with $\dd N/\dd t$ obtained from Eqs.~\eqref{eq:nimp} and~\eqref{eq:lossToN0}.
Equation~\eqref{eq:NdotT} then illustrates how each mode contributes to the depletion of the system. Particles initially present in the system (or incoming from a reservoir) with a distribution $n_{0,k}$ impinge on the loss barrier with a  velocity $v_k$ and are lost from the system with a probability $\eta_k$. We note in passing that Eq.~\eqref{eq:NdotT} takes the form of a Landauer formula for the loss current~\cite{Kamenev_book,Lebrat2019, Corman2019}.

From Eq.~\eqref{eq:nkt} it becomes evident that, in the case of a $T=0$ initial state, the discontinuity of the distribution at the initial Fermi surface persists at finite times at the same value of $k_F$. The presence of this discontinuity thus rationalizes the emergence of $\pi / k_F$-periodic Friedel oscillations in the density profile (cf. Eq.~\eqref{eq:FriedelOsc} and Sec.~\ref{sec:timeregimes}).

\begin{figure}[t!]
\centering
\includegraphics[width = 0.47 \textwidth]{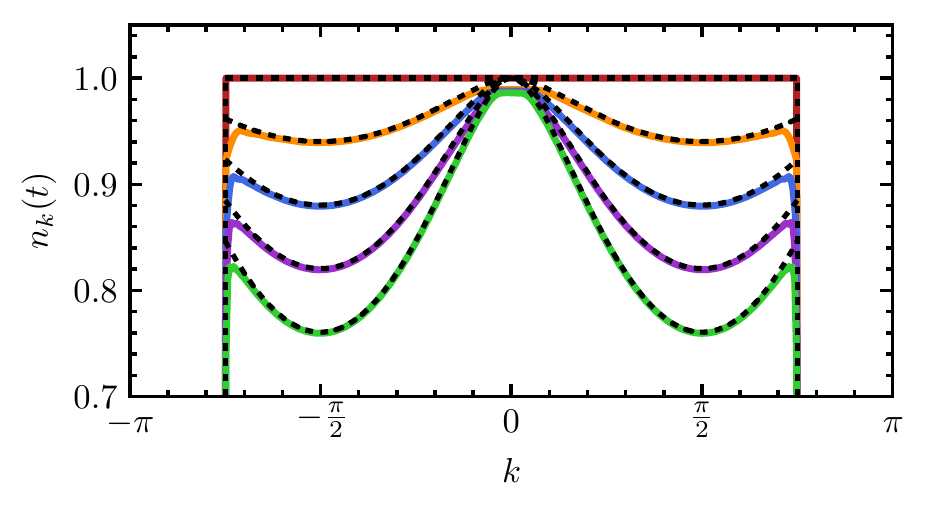}
\includegraphics[width = 0.47 \textwidth]{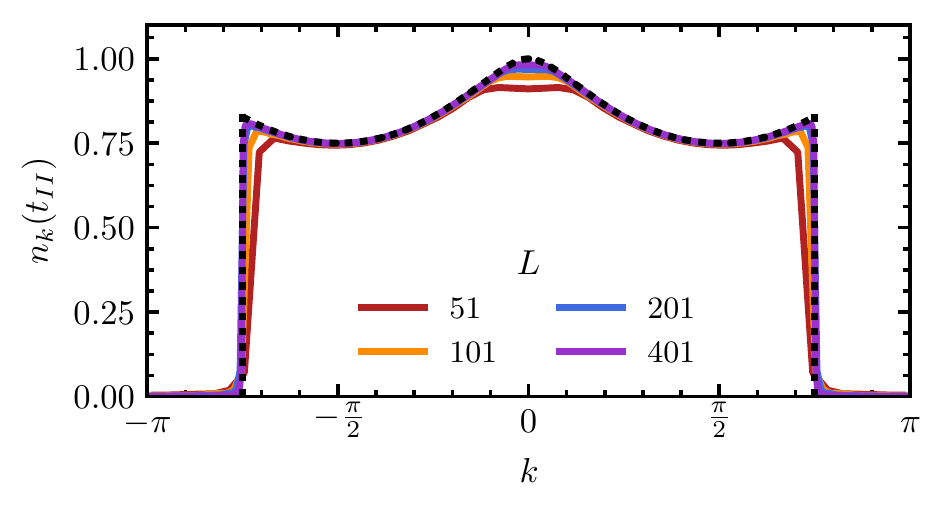}
\caption{%
(Color online). Momentum distribution $n_k(t)$ for the lattice model after the quench with solid lines indicating numerical simulations and dashed lines the analytical results. Upper panel: Momentum distribution in the second temporal regime for different times elapsed from the quench, $tJ = 0,  25,  50,  75, 100$. For all curves $L=401, \gamma = 3 J, N(0)/L=0.75, T/J=0$. Lower panel: Momentum distribution at the end of the second temporal regime for different system sizes, for a fixed aspect ratio $t_\text{II}= L/ (2 v)$. For all curves, $ \gamma = 2 J, N(0)/L=0.75, T/J=0$. For $N(0)/L=0.75$, $v= 2J$.
}
\label{fig:FermiNumerical} 
\end{figure}

Equation~\eqref{eq:nkt} can be adapted straightforwardly for the lattice model by using the appropriate $v_k$ and $\eta_k$. This allows us to further benchmark Eq.~\eqref{eq:nkt} with the full numerical solution for the lattice model~\eqref{eq:Hamiltonian-lattice} and to explore finite-size deviations. The comparison to the numerical simulations for a system prepared at $T=0$ is shown in Fig.~\ref{fig:FermiNumerical} (upper panel), indicating the depletion of $n_k(t)$ with a constant rate in the presence of a localized loss (solid lines).
A very good agreement with Eq.~\eqref{eq:nkt} (dashed lines) is found, except for small discrepancies at $k \approx 0$ and $k \approx \pm k_F$, due to finite-size effects.

In order to quantify these finite-size effects, we considered  the same momentum distribution for different system sizes; in each case we rescale the time elapsed from the quench as $t =L/2 v$, with $v$ the light-cone velocity, corresponding to the end of the second temporal regime $t_\text{II}$, thus rendering Eq.~\eqref{eq:nkt} independent of $L$. The results are displayed in Fig.~\ref{fig:FermiNumerical} (lower panel) and show the momentum distribution approaching the value predicted by Eq.~\eqref{eq:nkt} upon increasing the system size.

\subsubsection{Momentum distribution for the NESS}
\label{sec:nk_stationary} 

The momentum distribution of the NESS, $n_\text{ness}(k)$, can be computed by taking the Fourier transform of Eq.~\eqref{eq:correlationsfree} in the limit of $L,t \to \infty$, resulting in
\begin{equation}
n_\text{ness}(k) = n_{0,k} \left[ 
1 - \frac{ \eta_k}{2}
\right].
\label{eq:nkstationary} 
\end{equation}
As mentioned above, Eq.~\eqref{eq:nkstationary} cannot be simply obtained as a limit of Eq.~\eqref{eq:nkt}: in fact, it describes the momentum distribution \emph{inside} the light cone, where the system reaches a NESS continuously replenished by the currents originating from outside the light cone.
The factor $1/2$ in Eq.~\eqref{eq:nkstationary} comes from the fact that only half of the particles within the light cone have scattered off the impurity. This can be understood as follows: particles with momentum $k>0$, i.e., traveling rightward, are initially uniformly distributed along the wire. When the impurity is switched on, only the fraction of these particles at the left of the impurity will scatter on it, while the fraction on its right side will not. 
Therefore, only half of the particles with momentum $k>0$ will experience losses. The same argument holds true for particles with $k<0$.

\begin{figure}[t!]
\centering
\includegraphics[width = 0.47 \textwidth]{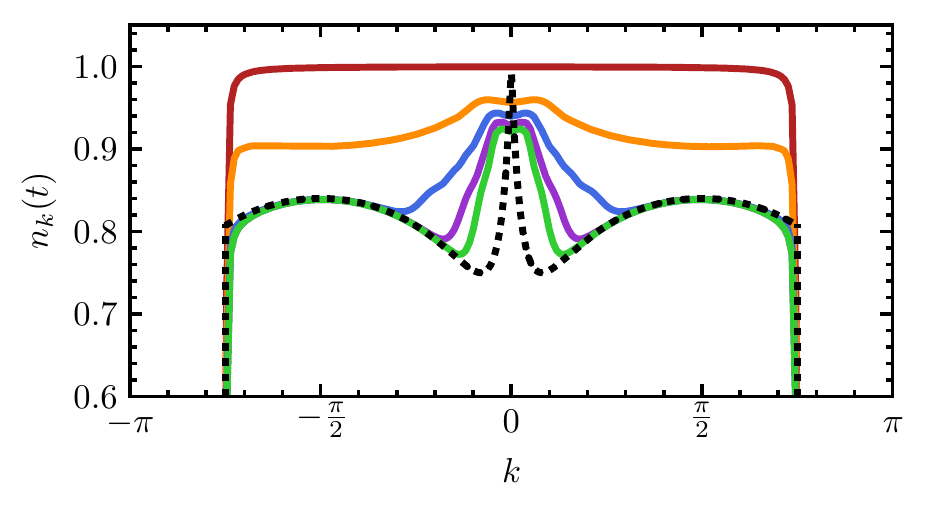}
\includegraphics[width = 0.47 \textwidth]{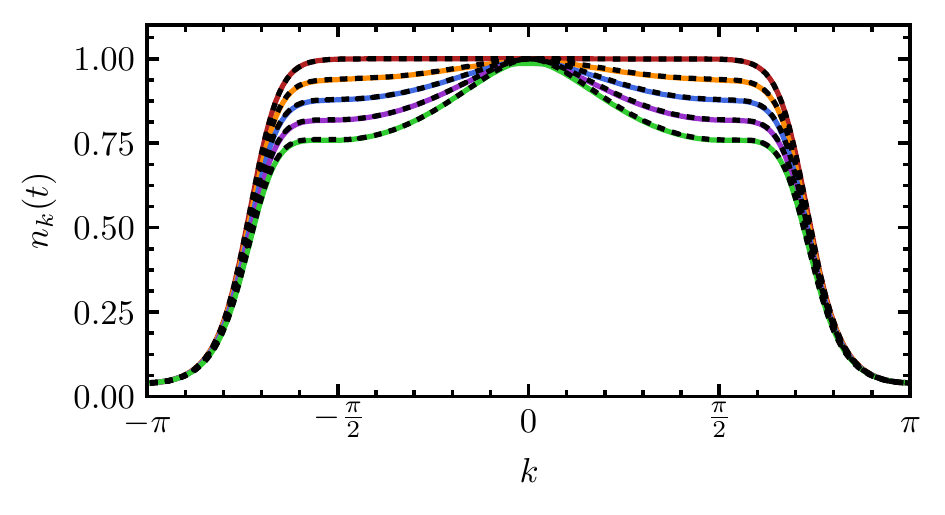}
\caption{%
(Color online). 
Upper panel: Momentum distribution $n_k(t)$ from the numerical simulations (solid) in the lattice model approaching a stationary value within the spatial segment $j\in[-100,100]$ of a system with $L=501$ sites, for different times elapsed from the quench in the second temporal regime, $t J = 0, 30, 60, 90, 120$. The dashed line, indicating the stationary distribution $n_\text{ness}(k)$, is approached asymptotically in time. For all curves $\gamma = 0.5 J, N(0)/L = 0.75, T/J =0$.
Lower panel: Momentum distribution $n_k(t)$ from the numerical simulations (solid) in the lattice model initialized with a finite temperature $T/J=0.2$ for different times after the quench in the second temporal regime, $t J = 0, 25, 50, 75, 100$. The dashed lines indicate analytical results. For all curves $L=401, \gamma = 3 J, N(0)/L=0.75$.
}
\label{fig:nkSubsectionFiniteT} 
\end{figure}

In a finite-size system, Eq.~\eqref{eq:nkstationary} is expected to describe the momentum distribution of a segment of the system within the light cone, up to finite-size effects. Experimentally, this quantity could be obtained in a time-of-flight measurement where the parts of the system not belonging to this segment are shielded from the detectors.
We report in Fig.~\ref{fig:nkSubsectionFiniteT} (upper panel) the asymptotic approach of $n_k(t)$ to the stationary distribution $n_\text{ness}(k)$ for a segment of size $L_{s}$ centered around the loss site, obtained from the numerical simulations of the lattice model \eqref{eq:Hamiltonian-lattice}, where $n_k(t)$ is computed from the Fourier transform of the correlations \eqref{eq:correlation} in this segment. For the finite times shown, slow modes with $|v_k| t < L_s/2$ have not yet relaxed to their stationary value in the considered segment, while the occupation of sufficiently fast modes approaches accurately the value predicted by Eq.~\eqref{eq:nkstationary}. 

%
Finally, as a consistency check, the comparison of Eq.~\eqref{eq:nkstationary} with Eq.~\eqref{eq:nNESSeta} reveals that the average density of the NESS corresponding to $n_\text{bg}$ is indeed obtained by summing over the momentum distribution in the NESS $n_\text{ness}(k)$.

\subsection{Finite temperature}
The presence of a finite temperature of the initial state is unavoidable in any experimental platform. 
In order to account for these effects, we consider the system to be prepared initially in a finite-temperature state $n_{0,k}$ before switching on the localized loss. In this section, we consider the effect of a finite-temperature initial state on the dynamics of the momentum distribution and the total particle loss rate. In Sec.~\ref{sec:cuttoffScales}, we will further comment on the effect of temperatures in the presence of interactions.

In Fig.~\ref{fig:nkSubsectionFiniteT} (lower panel) the evolution of $n_k(t)$ in the presence of a localized loss is shown for a system prepared at a finite temperature $T$, obtained from the numerical simulations of the lattice model \eqref{eq:Hamiltonian-lattice}. The comparison to Eq.~\eqref{eq:nkt} (dashed lines) with a thermal $n_{0,k}$ indicates a very good agreement with the numerical curves (solid lines). Besides the smearing effect of temperature, already present in the initial state, the same effect as for $T=0$ is observed, namely the depletion of the momentum distribution due to the losses. 
Moreover, because of the smearing, we expect the Friedel oscillations in the density profile to be damped out at distances larger than the thermal wavelength corresponding to the initial temperature. 

\begin{figure}[t]
\centering
\includegraphics[width = 0.47 \textwidth]
{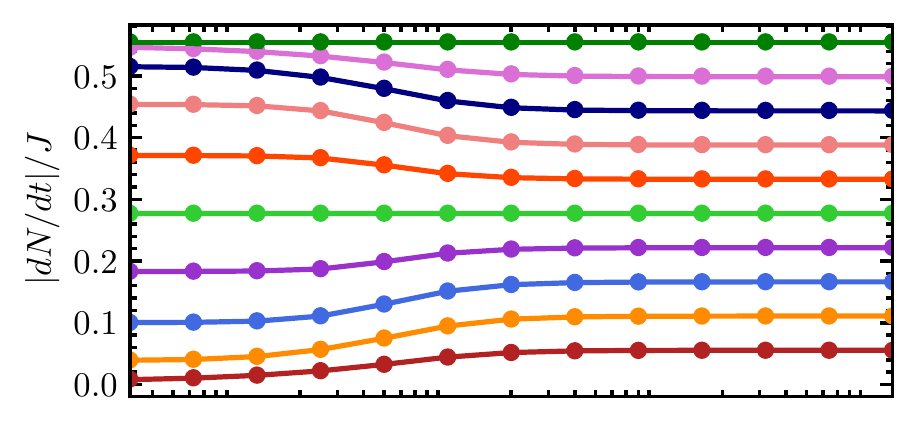}
\includegraphics[width = 0.47 \textwidth]
{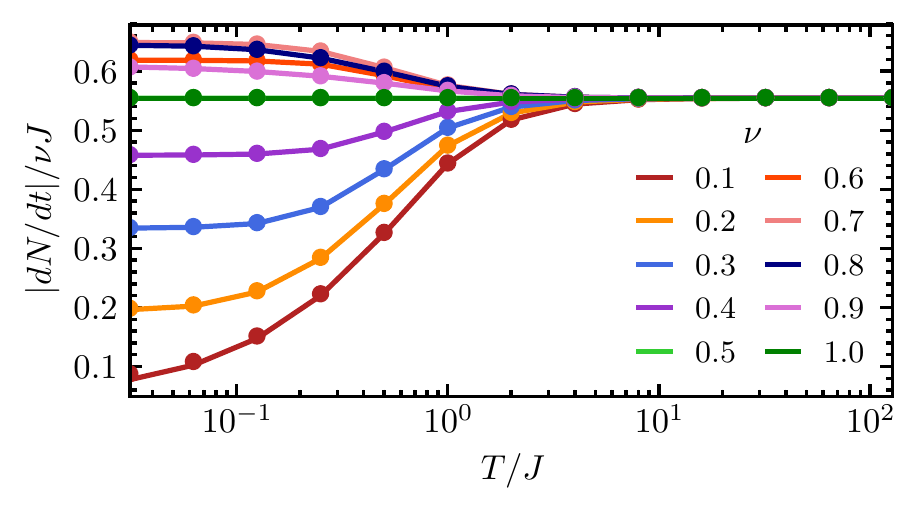}
\caption{%
(Color online). Particle loss rate $|dN/dt|$ (upper panel) and rescaled particle loss rate $|dN/dt|/\nu$ (lower panel) of the lattice model in the second temporal regime as a function of the temperature $T$ ($k_\text{B}=1$), for different initial filling factors $\nu = N(0)/L$. Dots indicate the results of numerical simulations, lines the analytical prediction.
For all curves $\gamma =3 J, L = 101$.
}
\label{fig:lossrate_T} 
\end{figure}

Further, we investigate how the interplay of temperature and filling of the initial state affects the total particle loss rate $|\dd N / \dd t|$ in the second temporal regime. In Fig.~\ref{fig:lossrate_T} (upper panel) we display  $|\dd N / \dd t|$ as a function of the temperature for different fillings, comparing the loss rates obtained from numerically simulating the lattice model \eqref{eq:Hamiltonian-lattice} (dots) with the analytical expression \eqref{eq:NdotT} (solid lines), indicating very good agreement. 
The loss rate exhibits a strong dependence on the initial particle filling. Below half filling, a higher temperature leads to a larger loss rate, while above half filling, a higher temperature leads to a lower loss rate. At half filling and at maximal filling, the loss rate is independent of the temperature. Moreover, for large temperatures $T$, the loss rate becomes independent of the temperature: in Fig.~\ref{fig:lossrate_T} (lower panel) the curves indicating the loss rate rescaled by the initial filling indeed collapse for large $T$.
A simple explanation for this behavior relies on the fact that the particle loss rate increases parametrically with the particle group velocity (cf. Eqs.~\eqref{eq:NdotT} and~\eqref{eq:defeta}), as faster particles scatter more frequently with the impurity. Since the group velocity on the lattice is nonmonotonic in $k$ (i.e., $v_k = 2J \sin k$), thermal fluctuation can increase or decrease the average particle group velocity, depending on the filling. In particular, the average particle velocity increases with the temperature below half filling (as states with higher group velocity are available to populate), while it decreases with the temperature above half filling (as only states with lower group velocity are available). When temperatures are high enough, the average particle velocity saturates to a finite value as every mode is occupied with the same probability, thus leading to a saturation of the particle loss rate as well.

\section{Real-space renormalization group approach}
\label{sec:realSpaceRG} 

In this section we consider the interplay between the interactions in the Hamiltonian~\eqref{eq:Hamiltonian-microscopic} and the presence of a localized loss.
The results will build on the analysis of the dynamics in the noninteracting system presented in Sec.~\ref{sec:freeSystem}. In particular, we will focus on interaction effects in the NESS for an infinitely long wire (see Sec.~\ref{sec:timeregimes}).
By making use of the properties of this NESS, we describe the effects of the interaction in terms of a renormalization of the scattering probabilities (cf. Sec.~\ref{sec:loss_probability}) based on a real-space RG approach initially developed for a coherent impurity at equilibrium~\cite{Matveev1993, Yue1994} and subsequently adapted to the case of a dissipative one~\cite{Froml2019}. 
This approach is perturbative in the microscopic interaction strength $V$ in Eq.~\eqref{eq:Hamiltonian-microscopic}, and valid for arbitrary dissipation strengths $\gamma$.
Due to its prominent role in the present analysis, we focus especially on the renormalization of the loss probability $\eta_k$, which is found to  vanish for modes near the initial Fermi momentum, for both repulsive and attractive interactions. The microscopic character of the approach allows for a transparent physical picture of these renormalization effects, 
which can be understood in terms of repeated virtual scattering processes between the bare loss barrier and an effective one created by the Friedel oscillations in the presence of interactions~\cite{Matveev1993, Yue1994}.
We then work out the observability of the renormalized loss probability in energy-resolved observables such as the energy-resolved loss rate and the momentum distribution of particles in the system.

We first derive the perturbative corrections to the scattering amplitudes $t_k, r_k$ due to the interaction strength $V$ for the continuum model~\eqref{eq:Hamiltonian-microscopic},
in analogy to the analysis of the potential barrier problem~\cite{Matveev1993, Yue1994}. 
These corrections will then be effectively resummed by RG flow equations for the scattering probabilities defined in Sec.~\ref{sec:loss_probability}.
The scattering problem can be conveniently formulated in terms of the retarded Green's function, as discussed in Sec.~\ref{sec:GreensMethod}. To this end, we consider corrections $\delta G$ to the retarded Green's function, $G +  \delta G$, where $G$ denotes the Green's function \eqref{eq:Gxx} in the presence of the localized loss but without interactions, i.e., $V=0$. 
The corrections can then be obtained within a first-order Born approximation as~\cite{Matveev1993, Yue1994,Froml2019}
\begin{multline}
\delta G(x,y,\omega) = \int_{x',y'} G(x,x',\omega)  \\ 
 \times \left[ V_H(x',y') + V_{ex}(x',y') \right] G(y',y,\omega)		,
\label{eq:GPert} 
\end{multline}
with the Hartree $V_H$ and exchange potentials $V_{ex}$ given by:
\begin{subequations}
\begin{align}
& V_H(x,y) = \delta(x-y)\int_{x'}V(x-x')C(x',x', 0), \\
& V_{ex}(x,y) = -V(x-y)\,C(y,x, 0) .
\end{align}
\end{subequations}
The correlation function $C(x,y,t)$ is evaluated in the stationary limit and reported in Eq.~\eqref{eq:correlationsfree}. We will consider the case in which the system is prepared at $T=0$, so that the single-particle correlations and density contain Friedel oscillations (see Eq.~\eqref{eq:FriedelOsc}). 
By evaluating Eq.~\eqref{eq:GPert} and by using the parametrization of $G$ in terms of $t_k$ and $r_k$ (see Eq.~\eqref{eq:GScattering}), in analogy to the case of a potential barrier~\cite{Yue1994, Matveev1993, Aristov2010}, the corrections to the scattering amplitudes for $k\sim k_F$ are obtained as
\begin{subequations}
\label{eq:perturbativecorrections} 
\begin{align}
\delta t_k & =  \alpha\, t_k r_k^2 \, \log|d(k-k_F)| ,\\
\delta r_k & =  \frac{\alpha}{2} r_k \bigg(r_k^2 + t_k^2 - 1 \bigg) \, \log|d(k-k_F)|.
\end{align}
\end{subequations}
Here, $d$ is a characteristic length scale which has to be chosen as the largest between the spatial range of the interaction $V(x)$ and the Fermi wavelength \cite{Fisher_review}.
The perturbative expansion is controlled by  $\alpha \ll 1$, with $\alpha = [\widetilde{V}(0)-\widetilde{V}(2k_F)]/(2\pi v_F)$,  with $v_F = k_F/m$ the Fermi velocity, and $\widetilde{V}(k)$ the Fourier transform of $V(x)$. Here, $\alpha>0$ corresponds to repulsive interactions, while $\alpha<0$ to attractive ones. 
The perturbative corrections can be interpreted  in terms of repeated virtual scattering processes between the bare barrier and the effective one generated by the Friedel oscillations in the presence of interactions~\cite{Matveev1993}.

Equations~\eqref{eq:perturbativecorrections} show logarithmic divergences for momenta $k \sim k_F$, originating from the interference between Friedel oscillations (see Eq.~\eqref{eq:FriedelOsc}) and particles with comparable wave vector. As a consequence, the perturbative expansion breaks down at $k \sim k_F$. However, these logarithmic divergences can be resummed in an RG scheme in real or frequency space~\cite{Yue1994, Matveev1993,Froml2019} resulting in the flow equations
\begin{subequations}
\label{eq:RG-equationsCoeff}
\begin{align}
\frac{\dd t(\ell)}{\dd \ell} & = - \alpha \, t(\ell) r^2(\ell)	, \\
\frac{\dd r(\ell)}{\dd \ell} & = - \frac{\alpha}{2} \,  r(\ell) \bigg( t^2(\ell) + r^2(\ell) - 1 \bigg)	,
\end{align}
\end{subequations}	
with the flow to be stopped at $\ell = -\log| d(k - k_F) |$. We notice that, under the RG flow, the relation $t_k=1+r_k$ is preserved and $t(\ell),r(\ell)$ remain real-valued, so that one can effectively consider just one of the two equations.
As our main interest is the characterization of the loss probability $\eta_k$, we infer from Eqs.~\eqref{eq:RG-equationsCoeff} the RG equations for the scattering probabilities \eqref{eq:scatteringProb}:
\begin{subequations}
\label{eq:RG-equations}
\begin{align}
\frac{\dd \mathcal{T}}{\dd \ell}  &= - 2\alpha \, \mathcal{T} \mathcal{R} ,\\
\frac{\dd \mathcal{R}}{\dd \ell}  &= - \alpha\, \mathcal{R}\left(\mathcal{R} + \mathcal{T} - 1\right)		,
\end{align}
\end{subequations}
with $\mathcal{T}, \mathcal{R} \equiv \mathcal{T}_k, \mathcal{R}_k$ for  $k\sim k_F$.
The renormalization of $\eta \equiv \eta_k$, for  $k\sim k_F$, can then be obtained from these flow equations with the use of Eq.~\eqref{eq:etadef}.

The flow equations admit the followings stable fixed points:
\begin{subequations}
\label{eq:fixedpointsTR} 
\begin{align}
&\mathcal{T}^*=0,\quad \mathcal{R}^*=1,\quad \eta^*=0 ,\qquad  \alpha>0 ,  \\ 
&\mathcal{T}^*=1, \quad  \mathcal{R}^*=0,\quad  \eta^*=0 ,\qquad \alpha<0	.
\end{align}
\end{subequations}
The transport properties of the system at the fixed points, characterized by $\mathcal{T}^*, \mathcal{R}^*$, are analogous to the case of a potential barrier: repulsive interactions result in perfect reflection for modes near $k_F$, while conversely for attractive interactions perfect transmission is restored despite the presence of a barrier.
As a central result, the loss probability $\eta$ vanishes for $\ell \rightarrow \infty$ for both repulsive and attractive interactions, i.e., $\eta^* = 0$. This entails that the loss of modes with momenta $k \sim k_F$ is suppressed, effectively restoring unitarity at the Fermi level. 

Before further commenting on the solution of the RG equations, we notice that $\mathcal{T}(k)$, $\mathcal{R}(k)$, and $\eta(k)$ can be parametrized with the use of a single dimensionless function $\widetilde{\gamma}(k)$  and study the RG flow of $\widetilde{\gamma}$. This procedure is practicable as under the RG flow the continuity relation $t_k=1+r_k$ is preserved as well as $t(\ell),r(\ell)$ remain real-valued, thus reducing the independent variables. We choose to parametrize the scattering coefficients by replacing $\gamma / |v_k| \rightarrow \widetilde \gamma$ in Eqs.~\eqref{eq:scatteringCoeff} and \eqref{eq:defeta}, according to which $\widetilde{\gamma}$ takes the role of an effective dissipation strength.
The RG flow of $\widetilde{\gamma}$ can be determined from the flow equations of the scattering amplitudes in Eqs.~\eqref{eq:RG-equationsCoeff} and reads
\begin{equation}
\label{eq:RG-gamma}
\frac{\dd \widetilde{\gamma}}{\dd \ell} = \alpha \frac{\widetilde{\gamma}^2}{1+\widetilde{\gamma}}.
\end{equation}
This equation admits respectively one stable fixed point $\widetilde{\gamma}^*= \infty$ for $\alpha>0$ and $\widetilde{\gamma}^*= 0$ for $\alpha<0$, corresponding to the fixed points of $\mathcal{T}, \mathcal{R}$, $\eta$ in Eqs.~\eqref{eq:fixedpointsTR}. We will comment more on the interpretation of $\widetilde{\gamma}$ in Sec.~\ref{sec:RG-comparison}.

\begin{figure}[t]
\centering
\includegraphics[width = 0.47 \textwidth]{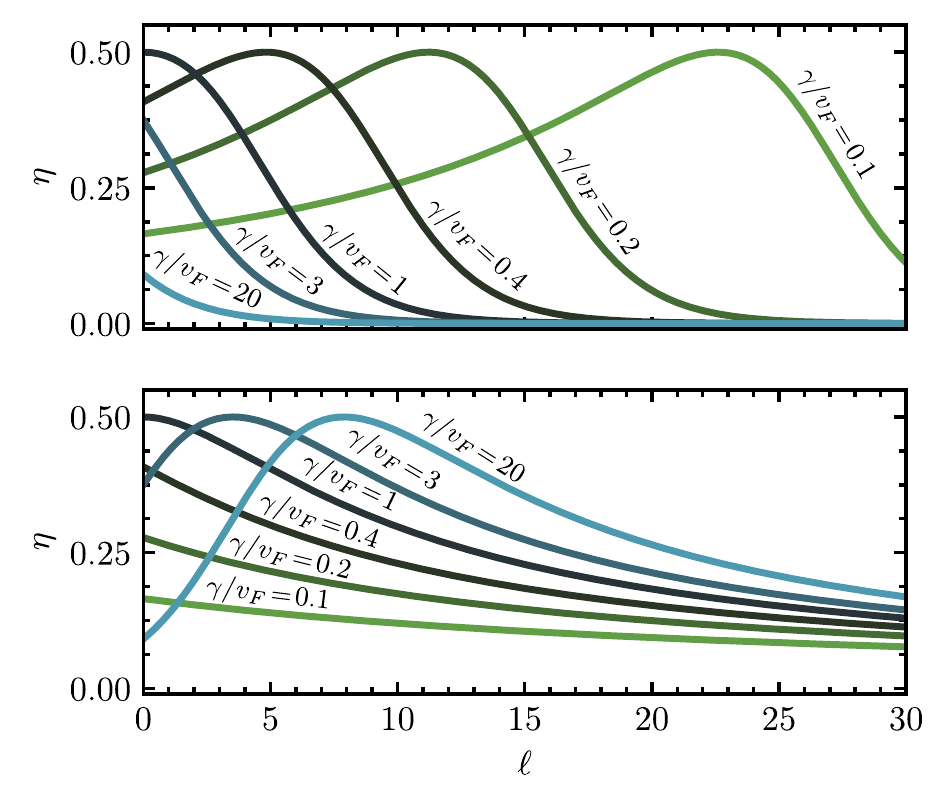}
\caption{%
(Color online). RG flow of the loss probability $\eta$ near the Fermi level for different bare values of $\gamma$. In the upper panel interactions act repulsively, i.e., $\alpha > 0$, in the lower panel attractively, i.e., $\alpha < 0$.
}
\label{fig:RGProb} 
\end{figure}

The RG flow of the loss probability $\eta$ is depicted in Fig.~\ref{fig:RGProb} for various dissipation strengths $\gamma$, where the latter determines the bare values from which the flow is initialized.
As an additional feature, $\eta$ can approach its fixed point value nonmonotonically (see Fig.~\ref{fig:RGProb}), in dependence on the bare value of $\gamma$ determining the initial conditions of the flow.
The asymptotic approach to the fixed points is qualitatively different from the equilibrium counterpart of a coherent potential barrier~\cite{Kane1992,Kane1992Long}. Focusing on $\eta$, we obtain in the vicinity of the fixed point, i.e., for $\ell \gg 1$, the asymptotic behavior
\begin{equation}
\label{eq:fluctuation-QZ}
\eta \sim 
\begin{cases}
 |k-k_F|^{\alpha} & \text{for} \quad \alpha > 0, \\
-1/\log|d(k-k_F)| & \text{for}\quad  \alpha < 0	.
\end{cases}
\end{equation}
The logarithmic approach for attractive interactions has no analog in the case of a coherent potential barrier, where the fixed points are approached algebraically in $|k-k_F|$ in both cases.

With use of the relation  $\ell = -\log| d(k - k_F) |$ one can reconstruct the renormalized value of $\eta$ for a given momentum in the vicinity of $k_F$ by stopping the RG flow at the corresponding scale. Consequently, the way the fixed point is approached (cf.  Fig.~\ref{fig:RGProb}) reflects onto the shape of $\eta_k$ for $k\sim k_F$.
For momenta close to $k_F$, $\eta$ always decreases, \emph{vanishing} exactly at $k_F$, for both repulsive and attractive interactions. However, depending on the bare value of $\gamma$, the RG flow can be nonmonotonic and thus the loss probability of modes on intermediate scales near $k_F$ can be enhanced before eventually dropping to zero at $k_F$: these modes thus experience a renormalized loss barrier resulting in an effectively enhanced loss probability.
 
\begin{figure}[t]
\centering
\includegraphics[width = 0.47 \textwidth]{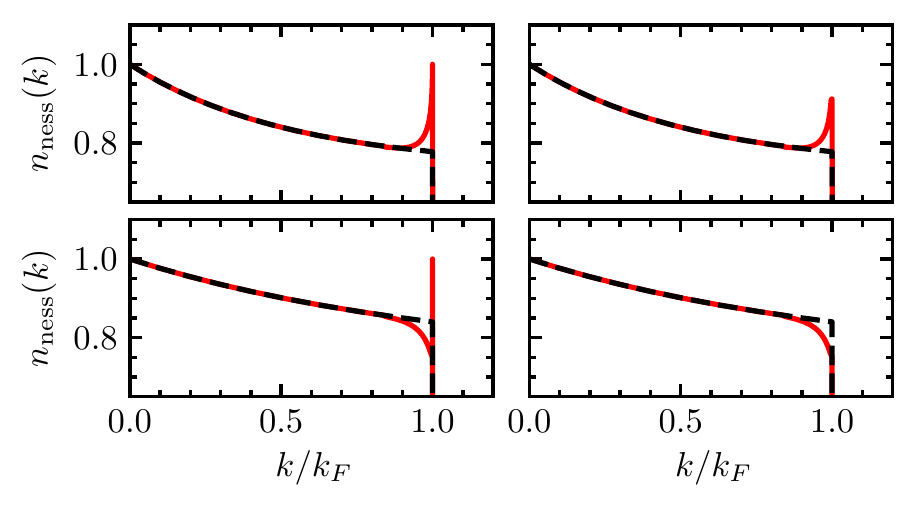}
\caption{%
Momentum distribution $n_\text{ness}(k)$  (solid red line) in the NESS, reconstructed from the RG flow of $\eta$. The momentum distribution of the noninteracting system (dashed black line) is shown for comparison.
The right panels show the same distribution as in the left panels but with a cutoff scale $\ell_{T/L} = 4$ (see Sec.~\ref{sec:cuttoffScales}). 
Upper panels: $\alpha = 0.5$ and $\gamma/v_F = 2$. Lower panels: $\alpha = -0.5$ and $\gamma/v_F = 4$.
}
\label{fig:RGEta} 
\end{figure}

As discussed in Sec.~\ref{sec:nK}, the momentum distribution $n_\text{ness}(k)$ of particles remaining in the system (cf. Eq.~\eqref{eq:nkstationary} and upper panel in Fig.~\ref{fig:nkSubsectionFiniteT}) bears a characteristic signature of the losses via a simple dependence on the loss probability $\eta$. The momentum distribution is expected to remain a good measure of the loss processes in the presence of interactions, as long as redistribution processes among the modes are small, i.e., for weak interactions, as described by the present approximation.
In Fig.~\ref{fig:RGEta} we display the momentum distribution $n_\text{ness}(k)$ as reconstructed from the RG flow of $\eta$, i.e., by replacing the bare $\eta_k$ by its renormalization-group improved counterpart.
The depletion is suppressed near $k_F$, resulting in a peak in the distribution of remaining particles. In the vicinity of $k_F$, either a domain of increased or suppressed depletion exists, in dependence on the bare parameters, as a consequence of the monotonic or nonmonotonic RG flow of $\eta$.

\section{Dynamical Hartree-Fock approximation}
\label{sec:HF}

In Sec.~\ref{sec:timeregimes} we investigated the dynamics of the noninteracting system in the presence of localized loss by making use of the correlation matrix $C_{ij} (t)$. The method can be straightforwardly adapted to include interactions within a Hartree-Fock approximation.
This allows us to study the effect of interactions on the results discussed in Sec.~\ref{sec:freeSystem}, such as the dynamical regimes, the particle loss rate, and the density profile after the quench.
In addition, we study how interactions modify the momentum distribution, and compare the result with the prediction discussed in Sec.~\ref{sec:realSpaceRG}.
We find indeed a suppression of losses near the Fermi momentum, resulting in an increased occupation of particles remaining in the system with momentum $k \sim k_F$ (cf.~Sec.~\ref{sec:nK}).

The dynamical Hartree-Fock approximation~\cite{Thouless2014quantum} is implemented as follows.
For the interacting Hamiltonian~\eqref{eq:Hamiltonian-lattice}, the equation of motion for the correlation matrix Eq.~\eqref{eq:correlationsDynamics} is not closed, as it depends on quartic correlators, which depend, in turn, on sextic operators and so on, resulting in a hierarchy of equations for higher order correlation functions.
The Hartree-Fock approximation consists then in closing the equations for the correlation matrix by reducing the quartic correlators into a product of quadratic correlators by the means of Wick's theorem.
The time evolution of the correlation matrix is then determined by Eq.~\eqref{eq:correlationsDynamics} with the following time-dependent non-Hermitian Hamiltonian:
\begin{align}
\label{eq:hf_hamiltonian} 
\widetilde H_\text{HF}(t)&= \sum_{j} \left[ \left( -J_{j}^{\vphantom{\dagger}}(t) \, \psi_{j+1}^\dagger \psi_{j}^{\vphantom{\dagger}} + \text{h.c.} \right)
 \right. \nonumber \\ & \hspace*{2cm} \left.
 - \mu_{j}^{\vphantom{\dagger}}(t) \, \psi_{j}^\dagger \psi_{j}^{\vphantom{\dagger}} - \ii \gamma \delta_{j,0}^{\vphantom{\dagger}} \psi_{j}^\dagger \psi_{j}^{\vphantom{\dagger}} \right]  	,
 \end{align}
where $J_j$ and $\mu_j$ are the effective hopping strength and on-site potential:
 \begin{subequations}
 \begin{align}
\label{eq:hf_mu} 
\mu_j(t) &=  -U \left[ C_{j-1,j-1}(t) + C_{j+1,j+1}(t) \right] , \\
\label{eq:hf_J} 
J_j(t) &= J + U C_{j,j+1}(t) 	.
\end{align}
 \end{subequations}
Since the initial state is translationally invariant, it is characterized by homogeneous values of $J_j$ and $\mu_j$, and described by Eqs.~\eqref{eq:correlationMatrixInitial} and \eqref{eq:S_initialCkk}  with a redefined hopping strength and chemical potential.
Observables are computed by solving numerically Eq.~\eqref{eq:correlationsDynamics} with the Hamiltonian~\eqref{eq:hf_hamiltonian} and initial state~\eqref{eq:correlationMatrixInitial}. 

\begin{figure}[t]
\centering
\includegraphics[width = 0.47 \textwidth]{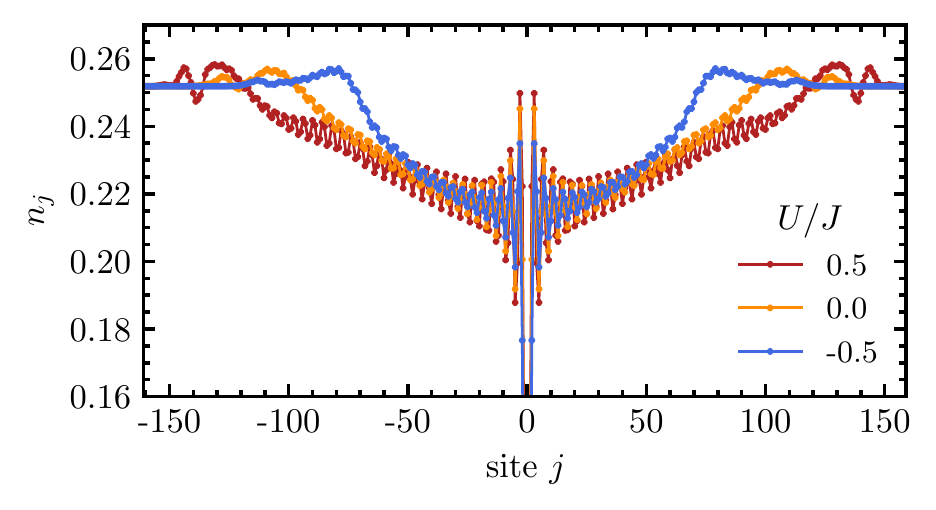}
\caption{%
(Color online). Density profile from the Hartree-Fock approximation for different interaction strengths at time $tJ=65$ elapsed from the quench. For all curves $\gamma=3 J, L=401, N(0)/L = 0.25$. 
}
\label{fig:hf_density} 
\end{figure}

In Fig.~\ref{fig:hf_density} the density profile is shown for repulsive ($U>0$, red curve) and attractive ($U<0$, blue curve) interactions, in comparison to the noninteracting system (yellow curve). The density profiles are evaluated after the same time elapsed from the quench for different interactions.
The density profile exhibits density oscillations which decay in space, with the same periodicity in all cases, determined solely by $k_F$ (cf.~Eq.~\eqref{eq:FriedelOsc}) as expected for Friedel oscillations also in the presence of interactions~\cite{Egger1995}. The amplitude of the oscillations on the other hand is increased (decreased) for repulsive (attractive) interactions.
The extent of the depleted region and consequently the propagation velocity of the light cone  (cf.~Sec.~\ref{sec:timeregimes}) is visibly dependent on the interaction strength. This can be understood by noticing that the group velocity $v_k$ (cf. discussion in Sec.~\ref{sec:timeregimes}) of the noninteracting problem is modified due to interactions. We compute these corrections by evaluating Eq.~\eqref{eq:hf_J} using the correlations of the initial state~\eqref{eq:correlationMatrixInitial} by which we obtain~\cite{Fowler1980}
\begin{equation}
\label{eq:hf_v} 
v_{\text{HF},k} = \left( 1+ \frac{U}{\pi J}  \sin k_F  \right)\, v_k	,
\end{equation}
i.e., the group velocity is modified by a factor depending only on the filling and interaction strength.
Hence, the propagation of the light cone is faster (slower) for repulsive (attractive) interactions as compared to the noninteracting system, in agreement with Fig.~\ref{fig:hf_density}.

\begin{figure}[t]
\centering
\includegraphics[width = 0.47 \textwidth]{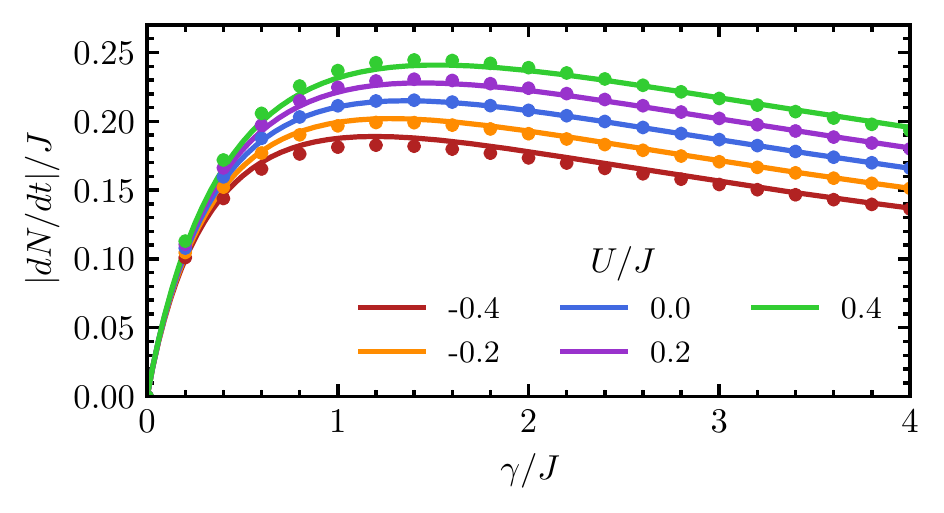}
\caption{%
(Color online). Total particle loss rate in the second temporal regime as a function of $\gamma$ for different interactions strengths $U$, for $L=151, N(0)/L = 0.4$ from the Hartree-Fock approximation (dots), in comparison to the analytical prediction (lines). 
}
\label{fig:hf_loss_rate} 
\end{figure}

The total particle loss rate in the second temporal regime (cf. Sec.~\ref{sec:timeregimes} and Fig.~\ref{fig:lossrate}) is shown in Fig.~\ref{fig:hf_loss_rate} as a function of $\gamma$ for different interaction strengths (dots). The nonmonotonic behavior due to the (microscopic) Zeno effect is clearly visible for all values of $U$. For all $\gamma$, the loss rate is increased for repulsive interactions while it is decreased for attractive ones. The dependence of the particle loss rate on the interaction strength can be described analytically (solid lines) by using the modified group velocity~\eqref{eq:hf_v}, resulting in good agreement with the numerical results.
More precisely, the solid lines represent analytical results obtained from Eq.~\eqref{eq:NdotT} with the interaction-dependent group velocity given by Eq.~\eqref{eq:hf_v}. Note that the group velocity enters also in the expression of the loss probability $\eta_k$~\eqref{eq:defeta}. 

According to the discussion in Sec.~\ref{sec:realSpaceRG}, we expect strong interaction effects on the loss probability near the Fermi momentum, suppressing the depletion of modes near $k_F$ for repulsive interactions. Since the total particle loss rate contains contributions of all momenta, it is not a well-suited observable for studying this effect. Instead, we consider the momentum distribution $n_k$, which can be regarded as a measure of the particles remaining in the system (cf. Sec.~\ref{sec:nK}). A suppression of loss at the Fermi edge reveals itself as an increased occupation in $n_k$ at these modes. 
In the presence of interactions $n_k$ remains a good measure of the loss processes as long as redistribution processes of the modes take place on timescales larger than the one of depletion processes. Within the present approximation, such redistribution processes are neglected.

\begin{figure}[t]
\centering
\includegraphics[width = 0.47 \textwidth]{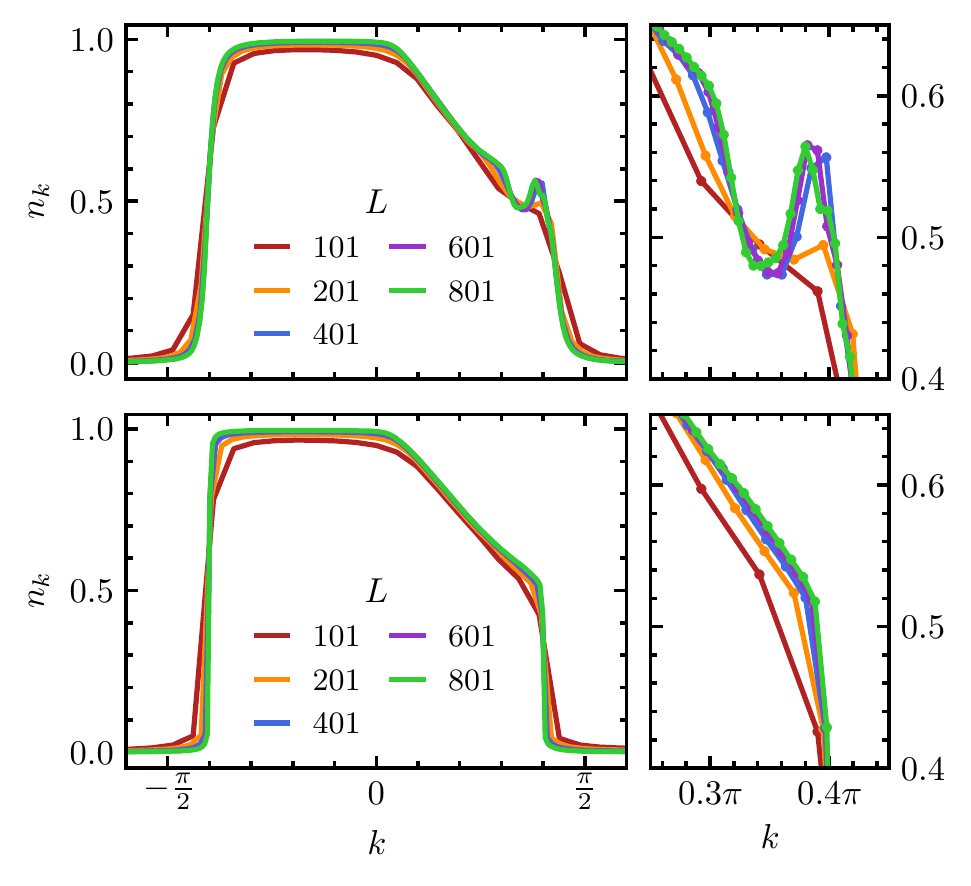}
\caption{%
(Color online). Momentum distribution $n_k$ for repulsive interactions $\nu U= 3/5 J$ (upper panels) and the noninteracting system $\nu U=0 J$ (lower panels) for increasing system sizes $L$. The distribution is obtained at $t J = L / 7$ close to the end of second temporal regime, in a segment with sites $j\in[0, 2 L /5]$ right to the impurity corresponding to the depleted region at that time. The smaller panels show a magnified view of the distribution in the respective panel to the left around $k_F = \nu \pi$.
For all curves $\gamma = 3 J, \nu = 0.4$.	}
\label{fig:hf_nk} 
\end{figure}

In Fig.~\ref{fig:hf_nk} (upper panels) we report the momentum distribution in a segment of the wire right to the loss site within the light cone (see discussion below and in Sec.~\ref{sec:nk_stationary}) at a time near the end of the second temporal regime. The momentum distribution reveals a pronounced peak in the occupation of modes at the Fermi momentum. As a comparison, in the noninteracting system such a peak is absent, see lower panels in Fig.~\ref{fig:hf_nk}. 
The presence of a peak at $k_F$ in the momentum distribution is consistent with the results of Sec.~\ref{sec:realSpaceRG} stating that the loss of particles with a momentum close to $k_F$ is reduced in the presence of interactions, resulting in an enhanced occupation at this scale.
The peak builds up with increasing system size, with which also the absolute size of the region exhibiting Friedel oscillations grows. The evaluation time is scaled with the system size such that the considered segment represents a constant fraction of the system.

The asymmetric shape of the momentum distribution in Fig.~\ref{fig:hf_nk} depends on the choice of a segment of the wire situated to the right of the loss site (but within the light cone).
With this choice, all particles with $k>0$ have scattered off the impurity and thus bear signature of depletion effects. Particles with $k<0$, instead, arrive from the right and have not encountered the loss site yet.
This specific segment is chosen in order to maximize the visibility of the peak, which is reduced by the smearing of the momentum distribution near the Fermi edge (see Figs.~\ref{fig:hf_nk} and~\ref{fig:FermiNumerical}). 
If a segment centered at the loss site is considered, the population of each momentum $k$ takes two contributions: from particles that do not encounter the loss (whose distribution resembles the left-hand side of the distribution in Fig.~\ref{fig:hf_nk}, upper-left panel), and from particles that do (whose distribution resembles the right-hand side of the distribution in Fig.~\ref{fig:hf_nk}, upper-left panel).
As a consequence, the peak would be averaged with the smeared, undepleted part of the distribution, thus lowering the visibility of the peak. 
We consequently can eliminate this contribution by choosing the segment only at one side of the impurity.
We also notice that the peak appears slightly to the left of $k_F$, which we interpret as a smearing effect.

We conclude with a brief discussion on the validity of the Hartree-Fock approach applied in the present section. Static Hartree-Fock approximation schemes have been applied to impurity and boundary problems in one-dimensional quantum systems~\cite{Cohen1998,Meden2000,Meden2002,Meden2003}.
Applications of self-consistent Hartree-Fock schemes have been discussed as leading in some cases to Friedel oscillations being nondecaying  at large distances from an impurity~\cite{Cohen1998}. This can be interpreted~\cite{Meden2003} as an onset of charge-density wave order different from the expected Luttinger liquid behavior.
We expect the dynamical Hartree-Fock approach applied here to capture qualitatively the local physics around the localized loss in the second temporal regime. An instability toward charge-density order would be signaled by a nondecaying amplitude of Friedel oscillations building up within the light cone. The study of charge-density ordered systems is itself an interesting problem where a fluctuation-induced QZE could emerge due the formation of a gap at $k_F$~\cite{Cohen1998}.

\section{Luttinger liquid description}
\label{sec:LuttingerLiquid} 

The Luttinger liquid is an effective theory able to capture the low-energy, collective behavior of gapless one-dimensional quantum systems with arbitrary interaction strengths~\cite{Giamarchi_book,Gogolin_book}. Among its many applications, it has been widely used in the study of impurities in fermionic systems~\cite{Gogolin_book}, notably the paradigmatic Kane-Fisher problem~\cite{Kane1992, Kane1992Long}.
In this section we describe the effects of the localized loss on a Luttinger liquid. We recall and extend the analysis performed in Ref.~\onlinecite{Froml2019}, up to second order in the impurity strength.

\subsection{Bosonization and Keldysh field theory}
As the system under consideration is driven out of equilibrium by the localized loss, a convenient functional description is provided by the Keldysh field theory~\cite{Kamenev_book,Sieberer_review}. After applying the bosonization transformations on the Hamiltonian and Lindblad operators, the quantum master equation~\eqref{eq:master-equation} can be incorporated into the Keldysh action~\cite{Mitra2011,Schiro2015, Buchhold2015}.
The interacting Hamiltonian~\eqref{eq:Hamiltonian-microscopic} hereby maps to the Luttinger liquid Hamiltonian~\cite{Giamarchi_book,Gogolin_book}
\begin{equation}
\label{eq:H-Luttinger}
H =  \frac{v}{2\pi} \int_x \left[ g\, (\partial_x \phi)^2 +g^{-1}\, (\partial_x \theta)^2  \right],
\end{equation}
with $\theta$ and $\phi$ real bosonic fields related to density and phase fluctuations, respectively, $v>0$ the sound velocity, and $g$ the so-called Luttinger parameter, both encoding the effect of interactions. In particular, $g<1$ corresponds to microscopic repulsive interactions and $g>1$ to attractive ones, with $g=1$ for a noninteracting system.
The low-energy physics of the models defined in Sec. II is captured for arbitrary interaction strength by the Luttinger Hamiltonian~\eqref{eq:H-Luttinger},  with the only exception being the lattice model at half filling which requires additionally $|U|<2$~\cite{Giamarchi_book}. In the latter case, the  Luttinger parameter $g$ can be expressed in a simple way as a function of the microscopic parameters, $g^{-1} = 2 \arccos(-U/2 J) /\pi$~\cite{Meden2002}.

The Keldysh action $S_0$ describing the thermal state $\propto \exp(-H/T) $, with $H$ given in Eq.~\eqref{eq:H-Luttinger}, can be written as
\begin{equation}
\label{eq:S0} 
S_0 = \frac{1}{2}\int_{k,\omega}\ \chi^\dagger (k, \omega) 
\begin{pmatrix}
0 & P_R^\dagger	(k, \omega)\\
 P_R(k, \omega) &  P_K(k, \omega)
\end{pmatrix}
 \chi(k, \omega),
\end{equation}
with $\int_{\omega} \equiv \int \dd \omega/(2\pi) $,  $\chi \equiv (\phi_c, \theta_c,  \phi_q,  \theta_q )^T$, where $c,q$ denote the classical and quantum fields~\cite{Kamenev_book}, and
\begin{subequations}
\label{eq:PR-PK}
\begin{align}
P_R &= \frac{1}{\pi} 
\begin{pmatrix}
-k^2 v g &  k(\omega +i\epsilon) \\
  k(\omega +i \epsilon) &   -k^2 v/g
\end{pmatrix}	
 , \\
P_K &=  \frac{2}{\pi} 
 \begin{pmatrix}
 0 & i  k \epsilon   \coth \frac{\omega}{2 T} \\
 i k \epsilon   \coth \frac{\omega}{2 T} &  0
 \end{pmatrix}. 
\end{align}
\end{subequations}
The infinitesimal dissipation $\epsilon$ acts as a regularization ensuring causality~\cite{Kamenev_book}, while the form of $P_K$ has been chosen in order to enforce the fluctuation-dissipation relation, and, consequently, thermal equilibrium~\cite{Sieberer_review}. 
The retarded and Keldysh Green's functions $G_R$ and $G_K$, respectively, are obtained as $G_R = P_R^{-1}$, and $G_K = - G_R P_K G_R^\dagger$.

The localized loss can be included in the Keldysh action by mapping the Lindblad superoperator of Eq.~\eqref{eq:master-equation} onto the action as~\cite{Sieberer_review}
\begin{equation}
\!\! S_\text{loss} = - i  \gamma \!\! \int_{x,t} \!\!\! \delta(x) \!\left[\LB_-^*  \LB_+ - \frac{1}{2} \left( \LB_+^* \LB_+ + \LB_-^* \LB_- \right)  \right]	\label{eq:Sdiss},
\end{equation}
with $\int_t \equiv \int_0^{+\infty}\dd t$, and $+,-$ the Keldysh contour indices. In turn, the fermionic Lindblad operators can be mapped onto bosonic degrees of freedom via
\begin{equation}
\label{eq:bosonization-mapping}
\psi \sim   e^{i k_F x}  e^{i (\phi + \theta)} +  e^{-i k_F x} e^{i (\phi - \theta)},
\end{equation} 
where only the two most relevant harmonics are retained~\cite{Kane1992Long,Giamarchi_book}.
By means of the mapping~\eqref{eq:bosonization-mapping} the impurity term~\eqref{eq:Sdiss} produces a backscattering ($S_\text{b}$) and a noiselike ($S_\text{n}$) contribution to the Keldysh action:
\begin{subequations}
\label{eq:Keldysh_impurity}
\begin{align}
\label{eq:backscattering}
S_\text{b} &= - 2i \gamma  \int_{x,t} \!\!\!\delta(x) \left( \ee^{i \sqrt{2}\phi_q} - \cos \sqrt{2}\theta_q \right)\cos \sqrt{2}\theta_c, \\
\label{eq:noise}
S_\text{n} &= - 2 i \gamma  \int_{x,t} \!\!\! \delta(x) \left( \ee^{\ii \sqrt{2}\phi_q} \cos\sqrt{2}\theta_q - 1 \right)	.
\end{align} 
\end{subequations}
The action thus acquires cosine terms, typical for impurity problems in a Luttinger liquid description. However, Eqs.~\eqref{eq:Keldysh_impurity} exhibit several differences with respect to the case of a coherent barrier~\cite{Kane1992,Kane1992Long}.
First, the dependence on $\phi_q$, absent for a coherent barrier, signals the breakdown of particle-number conservation due to the loss:  this term, in fact, explicitly breaks the global $U(1)$ symmetry of the system, which in the Keldysh bosonized action reads $\phi_c \to \phi_c + \alpha_c$, $\phi_q \to \phi_q + \alpha_q$, with $\alpha_{c,q}$ arbitrary real numbers \cite{Sieberer_review}. Moreover, the presence of odd powers of $\phi_q$ in the action entails that the average of the particle current is not zero, as expected. 

Second, the backscattering term~\eqref{eq:backscattering} differs from the one obtained from a coherent barrier. In particular, besides the prefactor being imaginary instead of real, the functional form differs from the one for a coherent barrier, i.e., $\sin \sqrt{2}\theta_q \sin\sqrt{2}\theta_c$~\cite{DallaTorre2012,Schiro2014}. This will lead to qualitatively different RG equations at second order, as shown in Sec.~\ref{sec:RG-luttinger}.
 
Third, the noise term $S_\text{n}$ in Eq.~\eqref{eq:noise} does not have a counterpart in the coherent case. As it directly descends from the quantum jump term in Eq.~\eqref{eq:master-equation} (i.e., the term containing $\LB^*_-\LB_+$ in Eq.~\eqref{eq:Sdiss}) and contains only Keldysh quantum fields, we conclude that it accounts for the quantum noise induced by the losses, and therefore it does not directly take part in the scattering induced by the impurity. Moreover, as discussed further below in Sec.~\ref{sec:RG-luttinger}, it is neither renormalized nor does it contribute to the renormalization of other terms on the action, at least up to second order in the impurity strength. For these reasons, we will not focus on it in the following.

\subsection{Weak and strong coupling: duality}
\label{sec:dual}

Given the nonlinearity of the impurity action~\eqref{eq:backscattering}, its effect can only be included perturbatively, which is possible for $\gamma \to 0$. However, in the limit $\gamma \to \infty$, the effect of the impurity is indistinguishable from the one of an infinitely strong barrier (cf. Eq.~\eqref{eq:romega}), namely it separates the wire into two semi-infinite, disconnected parts, with the density field pinned to zero at the impurity site~\cite{Giamarchi_book}. Accordingly, the effect of a very large but finite $\gamma$ can be treated as a perturbation around this disconnected-wire configuration.    

The perturbation is then implemented by a weak dissipative tunneling between the two parts of the wire, in analogy to the coherent tunneling for the case of a large but finite potential barrier~\cite{Kane1992, Kane1992Long, Fisher_review}.
The exact form of this dissipative tunneling is obtained by integrating out the dissipative site in leading order in $\gamma^{-1}$ as detailed in Appendix~\ref{sec:suppl:strongcoupling}, resulting in a tunneling term described by the Lindblad operator
\begin{equation}
\label{eq:diss_tunneling}
\LB= \psi_A(x=0) + \psi_B(x=0)		,
\end{equation}
with $\psi_{A,B}(x=0)$ the fermionic annihilation operators on the edges of the two semi-infinite wires. The dissipative tunneling strength is then quantified by a coupling $D \sim \gamma^{-1}$.
The bosonized representation of the electron annihilation operator of the weak link is given by $\psi_{A/B}(0) = \ee^{i \phi_{A/B}(0)}$, as the density fluctuations $\theta$ are frozen by the boundary constraint~\cite{Giamarchi_book}. It is then convenient to work with new fields $\varphi = \left(  \phi_{A} - \phi_{B} \right) / 2$, $\widetilde \varphi = \left(  \phi_{A} + \phi_{B} \right)  / 2 $, corresponding to symmetric and antisymmetric linear combinations of the fields in the two semi-infinite wires, respectively~\cite{Giamarchi_book,Gogolin_book}. This construction produces a local backscattering formally identical to $S_\text{diss}$ in Eq.~\eqref{eq:backscattering} upon the following replacements:
\begin{equation}
  \theta \rightarrow \varphi, \qquad \phi \rightarrow \widetilde \varphi, \qquad \gamma \rightarrow D.
\end{equation}
Moreover, the bare action~\eqref{eq:S0} exhibits a duality upon interchanging $\theta$ and $\phi$, and replacing $g$ with $g^{-1}$, which results in the correlation functions of phase and density fields being connected by a transformation $g~\to~g^{-1}$. For this reason, 
the action in the strong-coupling regime exhibits a duality with the one in the weak-coupling regime, resulting in the RG flow of $D$ being identical to the RG flow of $\gamma$ upon replacing $g$ with $g^{-1}$, as detailed in the next section.

\subsection{Renormalization group}
\label{sec:RG-luttinger}

The effect of the dissipative impurity on low-energy modes can be unveiled by studying its RG flow. This approach, pioneered for a coherent impurity in Refs.~\onlinecite{Kane1992,Kane1992Long}, has been shown~\cite{Froml2019} to capture at first order in the impurity strength the effects of the renormalization of the loss impurity consistently with the microscopic approach outlined in Sec.~\ref{sec:realSpaceRG}. 

However, in nonequilibrium Luttinger liquids, an effective temperature is typically generated during the RG flow, when nonlinear terms are included at second order in the RG equations~\cite{Mitra2006,Mitra2011,DallaTorre2012,Schiro2014}. These terms induce a coupling between high- and low-energy modes, resulting in a finite effective temperature emerging for the latter. Therefore, in order to investigate possible heating effects, we report in the following a second-order RG analysis for the loss impurity.

The RG procedure is carried out in the following way: The flow equations for the dissipation strength $\gamma$ and other couplings specified below are obtained by integrating out fast modes and subsequently restoring the original cutoff by a suitable rescaling of momenta and frequencies. The slow and fast mode separation is implemented as $\chi = \chi_\text{s} + \chi_\text{f}$ (with $\chi$ defined below Eq.~\eqref{eq:S0}), where $\chi_\text{s}$ is defined for momenta $|k| \in \left[ 0,\Lambda \ee^{-\ell} \right]$, while $\chi_\text{f}$ is defined on the momentum shell $|k| \in \left[ \Lambda \ee^{-\ell}, \Lambda \right]$, with $\Lambda$ an ultraviolet cutoff and $\ell>0$. The nonlinear term $S_\text{loss}$ is taken into account perturbatively, so that the integration over fast momenta can be performed by truncating the cumulant expansion to second order in $\gamma$ as
\begin{equation}
\label{eq:cumulants}
\langle \ee^{  i S_\text{loss} } \rangle_\text{f}  \simeq  \ee^{  i\langle  S_\text{loss}\rangle_\text{f} -\left( \langle S^2_\text{loss}\rangle_\text{f}-\langle S_\text{loss}\rangle_\text{f}^2\right)/2 },
\end{equation}
with the fast-mode average defined as
\begin{equation}
\langle \dots  \rangle_\text{f} = \int \mathcal{D} \chi_\text{f} \;   \dots  \,\ee^{iS_0[\chi_\text{f}]}.
\end{equation}
While the evaluation of $\langle S_\text{loss} \rangle_\text{f}$ is straightforward (see Ref.~\onlinecite{Froml2019} and Appendix~\ref{app:derivation-RG-Luttinger}) and only contributes to the renormalization of $\gamma$, the evaluation of $\langle S^2_\text{loss} \rangle_\text{f}$ is more involved, and it is detailed in Appendix~\ref{app:derivation-RG-Luttinger}. We emphasize here two important aspects emerging from this calculation: First, recalling that $S_\text{loss} = S_\text{b} + S_\text{n}$ (see Eq.~\eqref{eq:Keldysh_impurity}), it turns out that $S_\text{n}$ drops out of the RG equations, since it neither acquires any renormalization nor contributes to the renormalization of other couplings. Second, even if $S_\text{b}$ contains both fields $\phi$ and $\theta$, only correlations of $\theta$ enter the evaluation of the renormalization integrals: in this respect, this is analogous to the case of a coherent barrier, where only density fluctuations contribute to the renormalization~\cite{Kane1992,Kane1992Long,Schiro2014}.

Before presenting and discussing the RG equations, it is necessary to consider which couplings are expected to be renormalized, besides the impurity strength $\gamma$. While the global parameters $g$ and $v$ in Eq.~\eqref{eq:S0} cannot be renormalized from the localized loss, their local values at $x=0$ can acquire corrections. It is therefore convenient to work with the effective quadratic action for the density field at the loss site $\Theta \equiv \theta(x=0)$ (since $\phi$ does not enter the renormalization), which can be obtained by integrating out the fields at $x\neq 0$~\cite{Gogolin_book}, and gives
\begin{equation}
\label{eq:local_action}
S_0^{\Theta}=\frac{i}{\pi g } \int_{\omega}
\begin{pmatrix}
\Theta_c^* & \Theta_q^*
\end{pmatrix}
\begin{pmatrix}
0 & - \omega \\ 
\omega & 2 \omega \coth \frac{\omega}{2 T}
\end{pmatrix}
\begin{pmatrix}
\Theta_c \\
\Theta_q
\end{pmatrix}.
\end{equation}
This corresponds to the action of a quantum particle coupled to an Ohmic quantum bath imposing a temperature $T$ and friction $1/g$. 
%
%

The term $\omega \coth[\omega/(2 T)]$ in Eq.~\eqref{eq:local_action} is not suitable for being renormalized given the infinite series of powers entering the hyperbolic cotangent function, and therefore we replace it with a simpler function of $\omega$ interpolating between quantum and classical scaling, i.e., $|\omega| + 2T$~\cite{DallaTorre2012}.  Equation~\eqref{eq:local_action} can thus be rewritten as:
\begin{equation}
\label{eq:local_action_RG}
S_0^{\Theta}=\frac{i}{\pi} \int_{\omega}
\begin{pmatrix}
\Theta_c^* & \Theta_q^*
\end{pmatrix}
\begin{pmatrix}
0 & -\kappa \omega \\ 
\kappa \omega & 2\kappa_0|\omega| +4 \kappa T
\end{pmatrix}
\begin{pmatrix}
\Theta_c \\
\Theta_q
\end{pmatrix},
\end{equation}
where the prefactors of the terms $\omega \Theta^*_q\Theta_c$ (and c.c.) and $|\Theta_q|^2$ are expected to acquire contributions under renormalization and therefore are denoted as $\kappa$ and $T$, respectively. The microscopic value of $\kappa$ is given by $\kappa_0 = 1/g$. Notice that the term proportional to $|\omega|$ is expected to remain constant under renormalization, as perturbative corrections cannot generate a non-analytic function of the frequency, and therefore it remains proportional to $\kappa_0$. 

The same strategy can be used for the dual case discussed in Sec.~\ref{sec:dual}: in that case, the fluctuations of the field $\widetilde \varphi$ do not contribute to the renormalization, and they can be dropped out. Consequently, one can conveniently focus on the effective action for  $\Phi=\varphi(x=0)$ which is dual to the action in Eq.~\eqref{eq:local_action_RG} upon replacing $\Theta \to\Phi$ and $\kappa_0 = 1/g \to g$.

From Eqs.~\eqref{eq:Keldysh_impurity} and~\eqref{eq:local_action_RG} the bare dimensions of $\gamma$, $\kappa$, and $T$ can be inferred, leading to $\kappa \sim \Lambda^0$ and $\gamma~\sim~T~\sim~\Lambda$. It is then convenient to introduce the rescaled dimensionless quantities $\bar{\gamma} \equiv \gamma/\Lambda$ and $\bar{T} \equiv T/\Lambda$. The derivation of the RG equations for $\bar{\gamma}$, $\kappa$, and $\bar{T}$ is detailed in Appendix~\ref{app:derivation-RG-Luttinger} and we report here their final form:
\begin{subequations}
\label{eq:RG-bosonization}
\begin{align}
\frac{\dd \bar{\gamma}}{\dd \ell} 	& = \bar{\gamma} \left( 1 - \frac{\kappa_0}{\kappa^2} - 2 \frac{\bar{T}}{\kappa}\right)	,\\
\frac{ \dd  \bar{T} }{\dd \ell} 			& =  \bar{T}  +  \frac{\bar{\gamma}^2}{\kappa} I_{T} - \frac{\bar{\gamma}^2 \bar{T}}{\kappa}I_\kappa ,  \label{eq:RG_eq_temp}\\ 
\frac{\dd \kappa}{\dd \ell} 				& =  \bar{\gamma}^2 I_\kappa  ,
\end{align}
\end{subequations}
where the integrals $I_{T,\kappa} = I_{T,\kappa}(\kappa_0,\kappa,T) $ are reported in Appendix~\ref{app:derivation-RG-Luttinger}. In the following, we will study the flow of $\kappa$ and $T$ for the initial conditions $\kappa = \kappa_0$ and $T = 0$, which represent the microscopic model in a prequench state at zero temperature.
The solutions of Eqs.~\eqref{eq:RG-bosonization} will enable us to understand the emergence of an effective temperature $\bar{T}$ and its consequences for the flow of $\bar{\gamma}$.
We finally remark that exactly the same equations are obtained in the dual case discussed in Sec.~\ref{sec:dual}, upon replacing in Eqs.~\eqref{eq:RG-bosonization} $\gamma \to D$ and $\kappa_0 = 1/g \to g $. The solution of the RG equations is discussed in the following sections.

\subsection{First-order result}
\label{sec:RG-first-order}

Before we consider the full solution of Eqs.~\eqref{eq:RG-bosonization}, it is instructive to recall the solution of the RG equations to first order, i.e., to order $O(\gamma)$ ($O(D)$ in the dual case).
In this case, the flow of $T$ is trivial and admits an unstable fixed point at $T=0$, which is left for any finite initial temperature $T>0$, while $\kappa$ does not flow and its value is fixed by $\kappa_0$. If the initial temperature is tuned to $T=0$, $\gamma$ acquires the following renormalization: 
\begin{equation}
\label{eq:RG-luttinger}
\frac{\dd \bar \gamma}{\dd \ell} = (1-g) \bar \gamma.
\end{equation}
Remarkably, this RG equation is fully analogous to the flow equation obtained for the case of a weak potential barrier~\cite{Kane1992,Kane1992Long} despite the nonequilibrium nature of the present model.
The flow equation \eqref{eq:RG-luttinger} entails that for attractive interactions ($g>1$) the dissipation strength $\gamma$ is renormalized to zero by quantum fluctuations, being irrelevant in the RG sense. The loss barrier thus becomes fully transparent and losses of low-energy modes, i.e., near the Fermi energy, are suppressed. On the converse, for repulsive interactions ($g<1$) the dissipation strength $\gamma$ is infinitely enhanced by quantum fluctuations, and therefore it is a relevant perturbation in the RG sense.

\begin{figure}[t]
\centering
\includegraphics[width = 0.4 \textwidth]{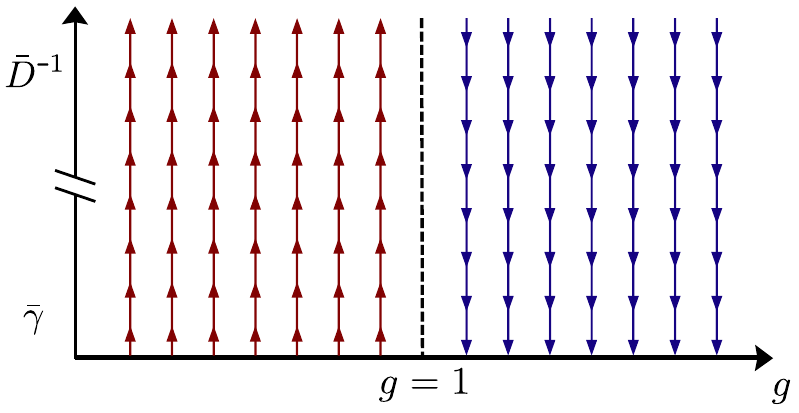}
\caption{%
Schematic RG flow of the dissipation strength $\gamma$, respectively $D$, in dependence on the Luttinger parameter $g$ at first order.}
\label{fig:rgschematics} 
\end{figure}

The RG flow equation for $\bar{D} \equiv \ D/\Lambda$ in the dual case reads: 
\begin{equation}
\label{eq:RG-luttingerStrong}
\frac{\dd \bar{D}}{\dd \ell} = (1-g^{-1}) \bar{D}.
\end{equation}
This result entails analogous behavior to the one obtained in the weak-coupling limit: for repulsive interactions, the weak link is cut, while for attractive interactions it is enhanced, reconnecting the two semi-infinite wires. The RG flow can thus be depicted as shown in Fig.~\ref{fig:rgschematics}: notice that the directions of the RG flows for $\gamma$ and $D$ are compatible. This leads to the conclusion that losses are always suppressed in the vicinity of the Fermi energy by gapless fluctuations, restoring unitarity at this scale, for both repulsive and attractive interactions.

\subsection{Second-order result}
\label{sec:RG_second_order}
The exact solution of Eq.~\eqref{eq:RG-bosonization} can only be accessed numerically, as the integrals $I_{\kappa, \bar{T}}$ cannot be solved in a closed form. Before turning to the solutions, we discuss some properties of the RG equations:
First of all, the integrals diverge for $g<1$ (resp. for $g>1$ for the dual case), implying that the perturbative analysis breaks down at $g =1$.  
This is tied to the fact that, for $g<1$ (resp. for $g>1$), $\bar{\gamma}$ (resp. $\bar{D}$) is a relevant perturbation. 
A similar breakdown of the perturbative analysis was pointed out for different nonequilibrium impurity problems in Refs.~\onlinecite{DallaTorre2012,Mitra2012,Schiro2014}.

Moreover, in the convergence region, the integrals $I_{T},I_\kappa$ are positive quantities, implying that $\bar{T}$ and $\kappa$ can only increase (see Fig.~\ref{fig:rg_integrals}). This is consistent with the expectation that a localized dissipative impurity {increases} the local effective temperature and the amount of friction (quantified by $\kappa$). Moreover, even if the initial RG value of the temperature is fixed to $T=0$, Eq.~\eqref{eq:RG_eq_temp} entails that a finite temperature is always generated. We emphasize that this temperature corresponds to a local effective temperature at the dissipative site, and not to a global temperature in the wire. 
\begin{figure}[t]
\centering
\includegraphics[width = 0.48 \textwidth]{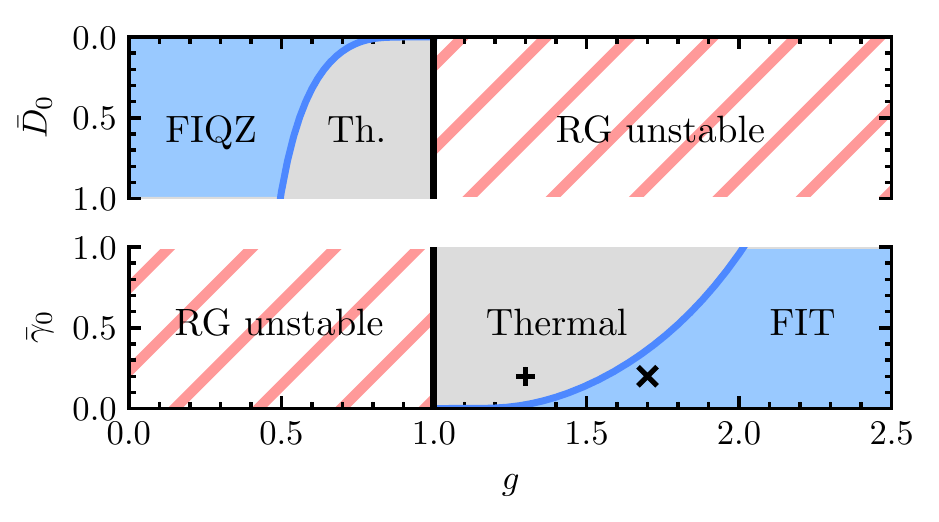}
\caption{%
(Color online). Phase diagram for the dissipative impurity from the second-order RG equations, for different values of the microscopic dissipation strength $\bar{\gamma}_0$ ($\bar{D}_0$ in the dual regime) and of the Luttinger parameter $g$. The blue-shaded area in the lower-right corner corresponds to values where the fluctuation-induced transparency is visible, while the upper-left one corresponds to values where the fluctuation-induced QZE is visible. The gray-shaded areas correspond to values where the effective temperature hinders these effects. The solid lines are obtained using Eqs.~\eqref{eq:c-scalte} and~\eqref{eq:T-scale} with $c=\log10$. The striped regions correspond to values where the perturbative RG breaks down.}
\label{fig:rg_phase_diagram} 
\end{figure}

The generation of a finite local effective temperature has the main effect of cutting off the RG flow of $\bar \gamma$, similarly to a finite global temperature in the pre-quench state or to a finite system size (see Sec.~\ref{sec:cuttoffScales}). In order to quantify this effect, a simple estimate can be carried out from Eqs.~\eqref{eq:RG-bosonization}. We define the RG scale $\ell_T$ as the scale at which the effective temperature $T$ becomes of the order of the UV cutoff $\Lambda$ or, equivalently, when the dimensionless temperature $\bar{T}$ becomes $\bar{T} \sim 1$. The regime of validity of the Luttinger description is then exited upon reaching this scale and, consequently, the RG flow has to be stopped. By focusing on the case $g>1$, when $\bar{\gamma}$ is irrelevant according to the first-order RG analysis (see Eq.~\eqref{eq:RG-luttinger}), we define a scale $\ell_{c}$ at which $\bar{\gamma}$ is decreased by a factor $\ee^{-c}$ from its microscopic value. Accordingly, for $\ell_c < \ell_T $, quantum fluctuations renormalize the value of $\bar{\gamma}$ by a factor $\ee^{-c}$ before the RG stops, thus unveiling the flow described in Sec.~\ref{sec:RG-first-order}. On the converse, if $\ell_c > \ell_T $, the RG flow stops before the quantum fluctuations renormalize $\bar{\gamma}$ by a factor $\ee^{-c}$. The value of $\ell_c$ can be estimated from Eq.~\eqref{eq:RG-luttinger} and reads:
\begin{equation}
\label{eq:c-scalte}
\ell_c = \frac{c}{g-1}.
\end{equation}
\begin{figure}[t]
\centering
\includegraphics[width = 0.48 \textwidth]{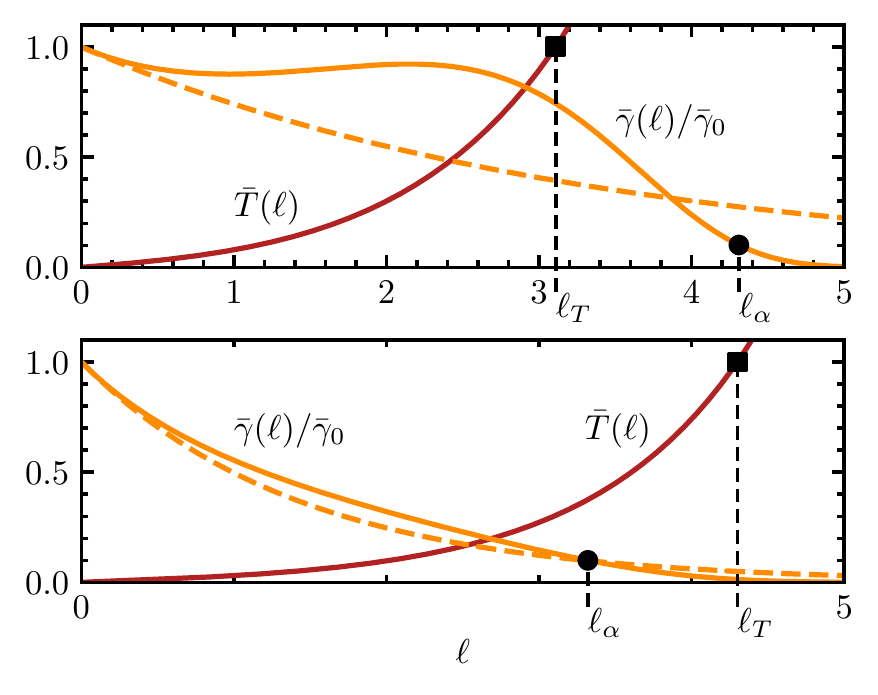}
\caption{%
(Color online). RG flow of $\bar{\gamma}$ and $\bar{T}$ from Eq.~\eqref{eq:RG-bosonization} with initial conditions $\bar{\gamma}(\ell=0)=\bar{\gamma}_0$ and $T(\ell=0)=0$. The dashed lines correspond to the solution of Eq.~\eqref{eq:RG-luttinger}. Upper (lower) panel: Values of $\bar{\gamma}_0$ and $g$ as indicated by the $+$ ($\times$) symbol in Fig.~\ref{fig:rg_phase_diagram}.}
\label{fig:rg_flow} 
\end{figure}
The scale $\ell_T$ can be estimated as follows. By replacing the solution of Eq.~\eqref{eq:RG-luttinger} into Eq.~\eqref{eq:RG_eq_temp} and solving to leading order in $\bar{\gamma}$ and for $\bar{T}(\ell=0)=0$, we thus obtain
\begin{equation}
\bar{T}(\ell) = \bar{\gamma}_0^2 \frac{g I_{0T}}{1-2g} \left( \ee^{2(1-g)\ell} - \ee^\ell \right) \approx \bar{\gamma}^2_0 \frac{g I_{0T}}{2g-1}  \ee^\ell 	,
\end{equation}
where $I_{0T} \equiv I_T(\kappa_0,\kappa_0,0)$, $\bar{\gamma}_0 \equiv \bar{\gamma}(\ell=0)$, and in the last equality we retained the growing exponential. From this equation one finds:
\begin{equation}
\label{eq:T-scale}
\ell_T = \log \left(\frac{2g-1}{\bar{\gamma}^2_0g I_{0T}}\right).
\end{equation}
The corresponding results in the dual regime can be obtained upon replacing $\bar \gamma \to \bar D$  and $g\to 1/g$.

The previous results are summarized in Fig.~\ref{fig:rg_phase_diagram} for $c = \log 10$: the blue-shaded area corresponds to values of $\bar{\gamma}_0$ and $g$ for which the scaling of Eq.~\eqref{eq:RG-luttinger} is visible ($\ell_c < \ell_T$), while the gray-shaded area corresponds to values of  $\bar{\gamma}_0$ and $g$ for which the effective temperature dominates over the scaling ($\ell_c > \ell_T$). The dashed areas correspond to regions where the perturbative analysis breaks down. The solid blue lines indicate the crossover between the two regions and are analytically determined from $\ell_c = \ell_T$. For weak interactions ($g \approx 1$), the effect of the effective temperature is dominant, overwriting the renormalization due to quantum fluctuations, which instead remain visible for larger interactions. Accordingly, the fluctuation-induced QZE and transparency are highly sensitive to self-thermalizing effects close to the critical point $g=1$.
In order to benchmark the previous estimates, we solve the full RG equations~\eqref{eq:RG-bosonization} numerically and compare them with the picture provided in Fig.~\ref{fig:rg_phase_diagram}. In Fig.~\ref{fig:rg_flow} (upper panel), the flow of $\bar{\gamma}$ and $\bar{T}$ is shown for values corresponding to the $+$ symbol in Fig.~\ref{fig:rg_phase_diagram}. The flow of $\bar{\gamma}$ is quite distorted compared to the scaling~\eqref{eq:RG-luttinger}, and decays only for $\ell >\ell_T$: the effect of quantum fluctuations is then washed out by the effective temperature, in agreement with the prediction in Fig.~\ref{fig:rg_phase_diagram}.
In Fig.~\ref{fig:rg_flow} (lower panel), instead, we show the flow of $\bar{\gamma}$ and $\bar{T}$  for values corresponding to the $\times$ symbol in Fig.~\ref{fig:rg_phase_diagram}. The value of $\bar{\gamma}$ almost reaches zero at $\ell_T$, leaving the scaling~\eqref{eq:RG-luttinger} largely visible, in agreement with the prediction drawn from Fig.~\ref{fig:rg_phase_diagram}.

\subsection{Relation to real-space RG approach}
\label{sec:RG-comparison}

A natural question concerns the relation between the real-space RG developed in Sec.~\ref{sec:realSpaceRG} and the RG developed in this section. While the two approaches were developed to encompass different regimes (arbitrary impurity strength and weak interaction in the former case, weak and strong impurity strength and arbitrary interaction for the latter case), a regime where both approaches are applicable exists (i.e., weak and strong impurity strength and weak interactions). Here, both approaches lead to the same result for the case of a coherent impurity~\cite{Fisher_review}. In this section, we investigate whether this connection persists for dissipative impurities.

This connection can be established by considering the parameter $\widetilde{\gamma}$ defined in Sec.~\ref{sec:realSpaceRG} (see above Eq.~\eqref{eq:RG-gamma}), which can be regarded as the renormalized strength of the impurity at $k \sim k_F$, in analogy to $\bar{\gamma}$ and $\bar{D}$ in the Luttinger liquid setting. 
For what concerns the Luttinger impurity strengths $\bar{\gamma}$ and $\bar{D}$, we will consider the first-order equations discussed in Sec.~\ref{sec:RG-first-order}, since the second-order equations account for coupling between different momenta, an effect not included in the real-space RG.

We study the flow equation for $\widetilde{\gamma}$ \eqref{eq:RG-gamma} in the limiting cases of weak and strong dissipation strength. In the limit $\widetilde D \equiv \widetilde\gamma^{-1}  \ll 1$, i.e., the strong-dissipation limit, we obtain 
\begin{equation}
\label{eq:LuttingerFlowEqStrong} 
\frac{\dd \widetilde D}{\dd \ell } =  - \alpha \widetilde D.
\end{equation}
This flow equation corresponds exactly to the one obtained in the Luttinger liquid RG in the strong-dissipation regime~\eqref{eq:RG-luttingerStrong} upon the identification $\widetilde D \equiv \bar{D}$, and by realizing that $\alpha$ relates to the Luttinger parameter $g$ as $\alpha \approx 1 - g \approx g^{-1} - 1$ for weak interactions ($\alpha \ll 1$, resp. $g \simeq 1$)~\cite{Fisher_review}. 
In the opposite limit of $\widetilde \gamma \ll 1$, we obtain the flow equation
\begin{equation}
\frac{\dd \widetilde \gamma}{\dd l } = \alpha \widetilde \gamma^2	.
\end{equation}
The comparison to the flow equation in the Luttinger liquid description $\eqref{eq:RG-luttinger}$ reveals that both equations indeed admit the same stable fixed points, however the approach to them differs.
With the microscopic analysis at hand, we can understand this discrepancy as follows: in the limit of large $\gamma$, the presence of the loss site is already taken into account in the unperturbed state (in the interactions) by the assumption of two decoupled systems. We thus perturb around a state in the second temporal regime, in agreement with the microscopic RG approach.
Compared with this, in the weak-coupling limit the unperturbed state is homogeneous and we operate rather in the first temporal regime. Nevertheless, the drastic renormalization effects obtained are consistent with the microscopic analysis, as Friedel oscillations build up also in the first temporal regime.

\subsection{Cutoff scales and experimental observability}
\label{sec:cuttoffScales} 
In general, the presence of a finite global temperature and system size both introduce a scale at which the RG flow is cut off. As discussed in Sec.~\ref{sec:RG_second_order}, also the effective temperature emerging at the impurity site can cut off the RG flow. 
In this section we provide estimates on the observability of the fluctuation-induced QZE when taking into account these scales.

\begin{figure}[t]
\centering
\includegraphics[width = 0.45 \textwidth]{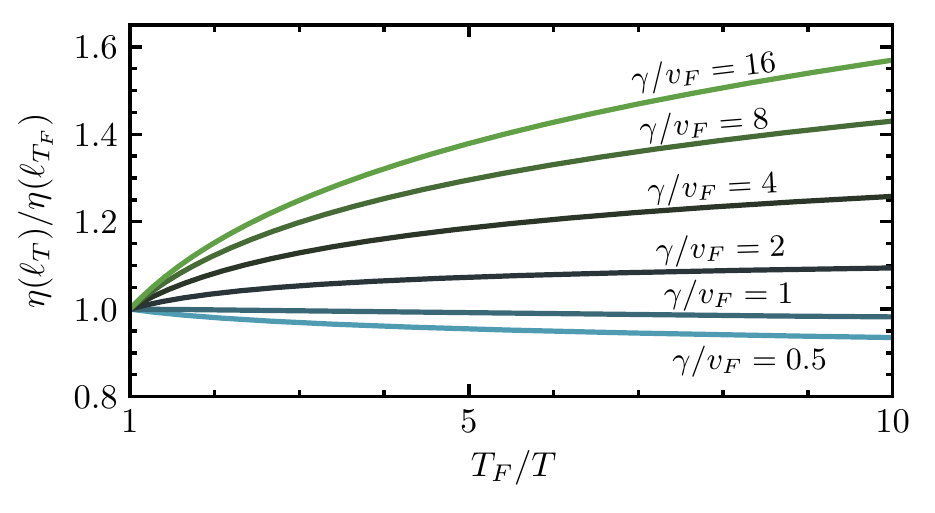}
\caption{%
(Color online).  $\eta(\ell_T)$ as a function of temperature $T$ for attractive interactions $\alpha=-0.5$ and different values of $\gamma / v_F$.}
\label{fig:experiments} 
\end{figure}

Let us first consider the effect of a finite global temperature $T$ and finite size $L$. A finite temperature cuts off the RG flow at a scale $\ell_T \simeq -\log(\sqrt{T/T_F})$, with $T_F$ the Fermi temperature. Analogously, a finite system size imposes a cutoff scale $\ell_L \simeq \log(L / d)$. Consequently, the variation of the temperature or system size allows one to probe the RG flow of $\eta(\ell)$ (cf. Fig.~\ref{fig:RGProb}), mapping out, for instance, the nonmonotonic behavior of the RG flow of $\eta$ (see Fig.~\ref{fig:RGEta}) as a function of $T$ or $L$.
In recent experiments ~\cite{Brantut_review,Lebrat2018,Lebrat2019,Corman2019} systems were realized that were typically characterized by sizes up to $L \sim 12 \mu m$ and temperatures down to $T \sim 0.1 T_F $, with $T_F$ the Fermi temperature, and Luttinger interaction parameters $g$ ranging between $1$ and $1.6$. The temperature scaling of $\eta$ using these exemplary parameters is shown in Fig.~\ref{fig:experiments}: an enhancement of $\eta$ up to 60\% from its microscopic value is expected in the range of temperatures considered, corresponding to the RG flow stopping at a scale $\ell_T \simeq 1.15$. 

Finally, we comment on the cutoff provided by the self-generated effective local temperature. While, in principle, such cutoff scale could be even more severe than the one provided by the global temperature and system size, we showed in Fig.~\ref{fig:rg_phase_diagram} that, upon choosing  proper microscopic values of the impurity strength $\gamma$ and of the interaction strength $g$, a regime can be achieved where the cutoff scale is sufficiently large to obtain a sizable renormalization.

\section{Conclusion and outlook}
\label{sec:conclusions}

In this work we discussed collective phenomena in an interacting one-dimensional ultracold gas of fermions driven out of equilibrium by the presence of a localized loss. We investigated the dynamics of the system after switching on the localized loss and identified the loss probability $\eta$ as the key quantity describing the loss properties of the system. We find that transmission and reflection of the wire are  drastically modified in the vicinity of the Fermi level, analogously to the paradigmatic case of a potential barrier. However, the RG scaling approaching the fixed point has no analog in the equilibrium counterpart. Moreover, the loss probability of momenta close to the Fermi momentum is found to be strongly renormalized, restoring unitarity at the Fermi level. These results can be interpreted as a fluctuation-induced QZE effect for repulsive interactions and a fluctuation-induced transparency for attractive ones.
Furthermore, we identified and characterized the visibility of this behavior in the momentum distribution $n(k)$ of the quantum wire, which is experimentally accessible. While in absence of interactions $n(k)$ exhibits a depletion profile due to the losses, the presence of interactions results in a suppression of losses close to $k_F$, resulting in an undepleted particle density at this momentum scale. 

While we mainly focused on fermions, our results are expected to apply also for one-dimensional Bose gases in the strong coupling regime, whose low-energy excitations also behave as a Luttinger liquid. In particular, transport across a localized potential barrier was shown to be strongly renormalized by fluctuations~\cite{Cominotti2014}, suggesting that the effects predicted in this work should also be visible there. A similar experimental setup was considered in Ref.~\onlinecite{Labouvie2016}.   

Additional future interesting directions involve the interplay of a localized dissipation and Luttinger physics on the nonequilibrium transport in ultracold fermionic wires coupled to imbalanced reservoirs~\cite{Lebrat2019,Corman2019}, as well as the interplay between dissipation and quantum interference when multiple dissipative impurities are included.

\begin{acknowledgments}
We thank I.~Carusotto, L.~Corman, H.~Ott, and M.~Schir\'{o} for useful discussions. We acknowledge support by the Institutional Strategy of the University of Cologne within the German Excellence Initiative (ZUK 81), by the funding from the European Research Council (ERC) under the Horizon 2020 research and innovation program, Grant Agreements No. 647434 (DOQS) and No. 648166 (Phonton), 
by the funding from the Deutsche Forschungsgemeinschaft (German Research Foundation) under Germany's Excellence Strategy – Cluster of Excellence 
Matter and Light for Quantum Computing (ML4Q) EXC 2004/1 – 390534769, 
by the Deutsche Forschungsgemeinschaft Collaborative Research Center (CRC) 1238 Project No. 277146847--projects C04 and C05, and CRC/Transregio 185 Project No. 277625399--project B3.
This research was supported in part by the National Science Foundation under Grant No. NSF PHY-1748958.
\end{acknowledgments}
 

\appendix

\section{Correlation matrix}
\label{sec:appendix_correlations} 

We consider the quantum master equation \eqref{eq:master-equation} on a lattice using the Hamiltonian \eqref{eq:Hamiltonian-lattice} with $U=0$. In the Heisenberg representation we obtain the equation for the evolution of the correlations $C_{ij}(t) = \langle \psi_i^\dagger(t) \psi_j(t) \rangle$ as
\begin{equation}
\frac{\dd}{\dd t}	C_{ij}(t) = \ii \sum_{k} \left( \widetilde H^\dagger_{ik} C_{kj}(t)  -  C_{ik}(t) \widetilde H_{kj} \right)	.
\label{eq:correlationsDynamics} 
\end{equation}
The dynamics of the time local correlations is then generated by the non-Hermitian effective Hamiltonian $\widetilde H$ given by
\begin{equation}
\widetilde H_{kl} =  -J \left(  \delta_{k,l+1} + \delta_{k+1, l} \right) - \ii \gamma \delta_{k,0} \delta_{l,0} 	.
\label{eq:Heff} 
\end{equation}
The solution of Eq.~\eqref{eq:correlationsDynamics} is reported in Eq.~\eqref{eq:correlation} of the main text, with the initial correlations given by
\begin{equation}
C_{j l}(0)
= \frac{1}{L^2} \sum_{k,k'} \ee^{-\ii k j} \ee^{\ii k' l} \langle\psi^\dagger_{k}(0) \psi_{k'}(0)\rangle,
\label{eq:correlationMatrixInitial} 
\end{equation}
where $k$ are the lattice momenta $k = 2 \pi n/L$, with $n \in \{j_\text{min}, j_\text{min} +1, \dots, j_\text{max}\}$ for periodic boundary conditions. The momentum correlation of the thermal homogeneous initial state $\langle\psi^\dagger_{k}(0) \psi_{k'}(0)\rangle$ is given by
\begin{equation}
\label{eq:S_initialCkk} 
\langle\psi^\dagger_{k}(0) \psi_{k'}(0)\rangle = \delta_{k, k'} \frac{1}{1 + \ee^{ ( \epsilon_k - \mu) / T}},		
\end{equation}
with $T$ the temperature, $\mu$ the chemical potential, and $\epsilon_k = -2J\cos k$ the lattice dispersion.

\section{Retarded Green's function}
\label{app:GreensFunction} 

In order to evaluate the retarded Green's function for the noninteracting continuum model, we map the quantum master equation  \eqref{eq:master-equation} onto a Keldysh action $S = S_0 + S_\text{loss}$~\cite{Sieberer_review, Kamenev_book}, with
\begin{equation}
S_0 = \int_{x,t}
\left[
i\psi_+^*\dot{\psi}_+ - H(\psi_+^*,\psi_+) - i\psi_-^*\dot{\psi}_- + H(\psi_-^*,\psi_-)
\right],
\end{equation}
and
\begin{equation}
\label{eq:S_Sloss}
S_\text{loss} = - i  \int_{x,t} \Gamma(x)  \left[\LB_-^*  \LB_+ - \frac{1}{2} \left( \LB_+^* \LB_+ + \LB_-^* \LB_- \right)  \right],
\end{equation}
with $\LB_\pm(x) = \psi_\pm(x)$.
We can evaluate the retarded Green's function of the noninteracting system with localized loss exactly using a Dyson equation~\cite{Kamenev_book}:  
\begin{equation}
\label{eq:DysonS}
G(x,x',\omega) = G_0(x,x',\omega) + \int_y G_0(x,y,\omega)\Sigma(y)G(y,x',\omega) .
\end{equation}
Here $G_0$ is the retarded Green's function of the homogeneous system, i.e., in the absence of the dissipative impurity:
\begin{equation}
\label{eq:G0xx} 
G_0(x,x',\omega) = \frac{\sqrt{2 m}}{2i\sqrt{\omega}} \ee^{i\sqrt{2 m \omega}|x-x'|}	.
\end{equation}
Due to the quadratic structure of $S_\text{loss}$ \eqref{eq:S_Sloss} the self-energy is field independent, and the Dyson equation can be written as
\begin{equation}
\!G(x,x',\omega) \!=\! G_0(x,x',\omega) -i\gamma G_0(x,0,\omega)G(0,x',\omega)		.
\end{equation}
From this expression, the solution of the retarded Green's functions is readily obtained as
\begin{align}
&G(x,x',\omega) 
= G_0(x,x',\omega) -i\gamma\frac{G_0(x,0,\omega)G_0(0,x',\omega)}{1+i\gamma G_0(0,0,\omega)} \nonumber \\
&= \frac{\sqrt{2m}}{2i\sqrt{\omega}} \left[ \ee^{i\sqrt{2m \omega}|x-x'|} + r(\omega)\,  \ee^{i\sqrt{2m \omega}(|x|+|x'|)} \right],
\label{eq:S_Gxx} 
\end{align}
where
\begin{equation}
r(\omega) = \frac{-\gamma}{\sqrt{2 \omega /m} + \gamma}		.
\label{eq:romegaS} 
\end{equation}
The interpretation as a scattering problem (cf.~Sec.~\ref{sec:loss_probability} in main text) can be strengthened by considering a mixed momentum and real-space representation
\begin{equation}
\label{eq:Gxq}
G(x,q,\omega) = \frac{1}{\omega+i \delta - \epsilon_q} \left[ \ee^{iqx} + r(\omega)\,  \ee^{i\sqrt{2m \omega}|x|} \right],
\end{equation}
where $\delta$ is an infinitesimal dissipation whose presence ensures causality. In fact, $G(x, q, \omega)$ corresponds to the response to an external perturbation $V= \int_x \left( h(x,t) \psi^\dagger{x} +\text{h.c.} \right)$,
with $h(x,t) = h_0 \exp{[\ii q x - \ii \omega t}]$, describing the injection of a free particle (hence described as a plane wave) incoming from the left, i.e., 
\begin{equation}
G(x,q,\omega) = \frac{\partial \langle \psi(x) \rangle_{h_0} }{\partial h_0}\biggr|_{h_0=0}.
\end{equation}
For $\omega = \epsilon_q$, Eq.~\eqref{eq:Gxq} has the asymptotic behavior as reported in Eq.~\eqref{eq:GScattering}.

\section{Correlation functions}
\label{sec:correlationFunctions} 

We derive here an exact expression for the two-time one-particle correlation functions
\begin{equation}
C(x,x',t,t') = \langle \psi^\dagger(x,t)\psi(x',t')\rangle.
\end{equation}
The system is assumed to be prepared in a homogeneous state characterized by a momentum distribution $n_{0,k}$, and the localized loss is switched on at time $t_0$. Since no particles are injected in the system, the dynamics of the one-particle correlation functions can be obtained as a propagation of the initial correlations as:
\begin{align}
\label{eq:Cxy-definition}
&C(x,x',t,t') \nonumber \\
&= \int_{y,y'} G^*(x,y,t-t_0)G(x',y',t'-t_0)  C(y,y',t_0,t_0) \nonumber \\
& = \int_k \, G^*(x,-k,t)G(x',-k,t') n_{0,k}	,
\end{align}
where we used the retarded Green's function $G(x,x',t) $ given in Eq.~\eqref{eq:Gxx}, and the initial correlations $C(x,x',t_0,t_0)$ given by
\begin{equation}
 C(x,x',t_0,t_0) = \int_k \ee^{ik(x'-x)} n_{0,k}.
\end{equation} 
The integrand in Eq.~\eqref{eq:Cxy-definition} can be evaluated as follows:
\begin{multline}
 G^*(x,-k,t)G(x',-k,t')  \\
 =  \int_{\omega, \omega'} \frac{\ee^{i\omega(t-t_0)} \ee^{-i\omega(t'-t_0)}  f^*(\omega,k,x) f(\omega',k, x')}{(\omega - i \delta - \epsilon_k) (\omega' + i \delta - \epsilon_k)}  	,
\end{multline}
where $f(\omega,k,x) = \ee^{-ikx} + r(\omega)\ee^{i\sqrt{2m \omega}|x|}$. Then, by making use of 
\begin{equation}
(\omega' + i \delta - \epsilon_q)^{-1} = -\ii \int_{t''}\, \theta(t'') \ee^{i(\omega' - \epsilon_q)t''}	,
\end{equation}
and of the identity
\begin{equation}
 \ee^{i\omega t_0} \int_t\, \theta(t) \ee^{i(\omega - \epsilon_k)t} =  \int_t \theta(t- t_0) \ee^{i(\omega - \epsilon_k)t} = 2 \pi \delta(\omega - \epsilon_k)	,
\end{equation}
where in the last equality we took the limit $t_0 \rightarrow -\infty$ in order to obtain the expression in the stationary regime, we arrive at Eq.~\eqref{eq:correlationsfree} of the main text.

\section{Lattice}
\label{sec:S_lattice} 
We present here the derivation of the retarded Green's functions for the noninteracting lattice Hamiltonian~\eqref{eq:Hamiltonian-lattice} with a dissipative loss at $j=0$. In analogy to the continuum case in Appendix~\ref{app:GreensFunction} we obtain the retarded Green's function in momentum space $G(k,k',\omega)$ as
\begin{equation}
\label{eq:Dyson-lattice-momentum}
G(k,k',\omega) = G_0(k,\omega) \delta_{k,k'} - \frac{i \gamma G_0(k,\omega) G_0(k',\omega) }{1 + i \gamma {G_0}_{j=0}(\omega)}	,
\end{equation}
where ${G_0}(k,\omega)$ is the retarded Green's function of the homogeneous system ($\gamma=0$) in momentum space and $ {G_0}_{j}(\omega)$ is the retarded Green's function of the homogeneous system in real space, which read, respectively
\begin{equation}
\label{eq:S_lattice_G0k} 
{G_0}(k,\omega) = (\omega + i \delta - \epsilon_k)^{-1} ,\qquad \epsilon_k = -2 J \cos(k),
\end{equation}
with $\delta \to 0^+$ an infinitesimal dissipation which guarantees causality, and, by taking the Fourier transform of~\eqref{eq:S_lattice_G0k},
\begin{equation}
\label{eq:retarded-lattice}
{G_0}_{i-j, \omega} = \frac{ \left( - \frac{\omega}{2 J} + \ii  \sqrt{1 - \frac{\omega^2} {4 J^2} } \right)^{{|i-j|} } } { 2 J \ii \sqrt{1-  \frac{\omega^2}{4 J^2} }}.
\end{equation}
From Eq.~\eqref{eq:Dyson-lattice-momentum}, we obtain the retarded Green's function in real space $G_{i,j}(\omega)$:
\begin{equation}
G_{i,j}(\omega) = {G_0}_{i - j}(\omega) - \frac{i \gamma {G_0}_i(\omega)  {G_0}_j (\omega) }{1+ i \gamma {G_0}_{j=0}(\omega)},
\end{equation}
which, using Eq.~\eqref{eq:retarded-lattice}, takes the explicit form of a scattering problem
\begin{equation}
G_{i,j}(\omega) = \frac{1}{ \ii \sqrt{4-\frac{\omega^2}{ J^2}}} \left[ \ee^{\ii f(\omega) |i - j|} + r(\omega) \ee^{\ii f(\omega)( |i| - |j|)} \right],
\end{equation}
with
\begin{equation}
f(\omega) = - \ii \ln \left(-\frac{\omega}{2 J} + \ii \sqrt{1- \frac{\omega^2}{4 J^2}} \right),	
\end{equation}
and
\begin{equation}
r(\omega) = - \frac{  \gamma  }{
v(\omega) + \gamma  }	,
\label{eq:c_rsrg:reflection_compact} 
\end{equation}
with the group velocity $v(\omega) = J \sqrt{4-\omega^2 / J^2}$. From this, we can identify $r(\omega)$ as the reflection amplitude and $t = 1+r$ as the transmission amplitudes.
In a mixed real-space and momentum representation, we obtain
\begin{equation}
G_j (k, \omega) = G_0(k,\omega)  \left[ \ee^{ \ii k j} + r(\omega) \ee^{\ii f(\omega) | j |} \right].
\end{equation}
As discussed in Appendix~\ref{app:GreensFunction}, we can use the mixed representation to quantify the response to a plane wave perturbation with frequency $\omega = \epsilon_k$. In this case, we obtain the concise expression $f(\omega) = |k|$, in analogy to the continuum case~\eqref{eq:Gxq}. 
The similar form of the lattice Green's function and scattering coefficients in comparison to the continuum case allows one to adapt the results of Sec.~\ref{sec:freeSystem} straightforwardly by replacing the dispersion relation with the one of the lattice and integrations with their lattice counterparts.

\section{Spectral properties of the Green's function and localized state on a lattice}
\label{sec:S_lattice_bound_state} 

In this appendix the spectral properties of the retarded Green's function and of the associated non-Hermitian Hamiltonian are discussed for a noninteracting system on a lattice. We show that the spectrum exhibits a transition, with a localized state emerging above a critical value of $\gamma$.

\begin{figure}[t!]
\centering
\includegraphics[width = 0.45 \textwidth]{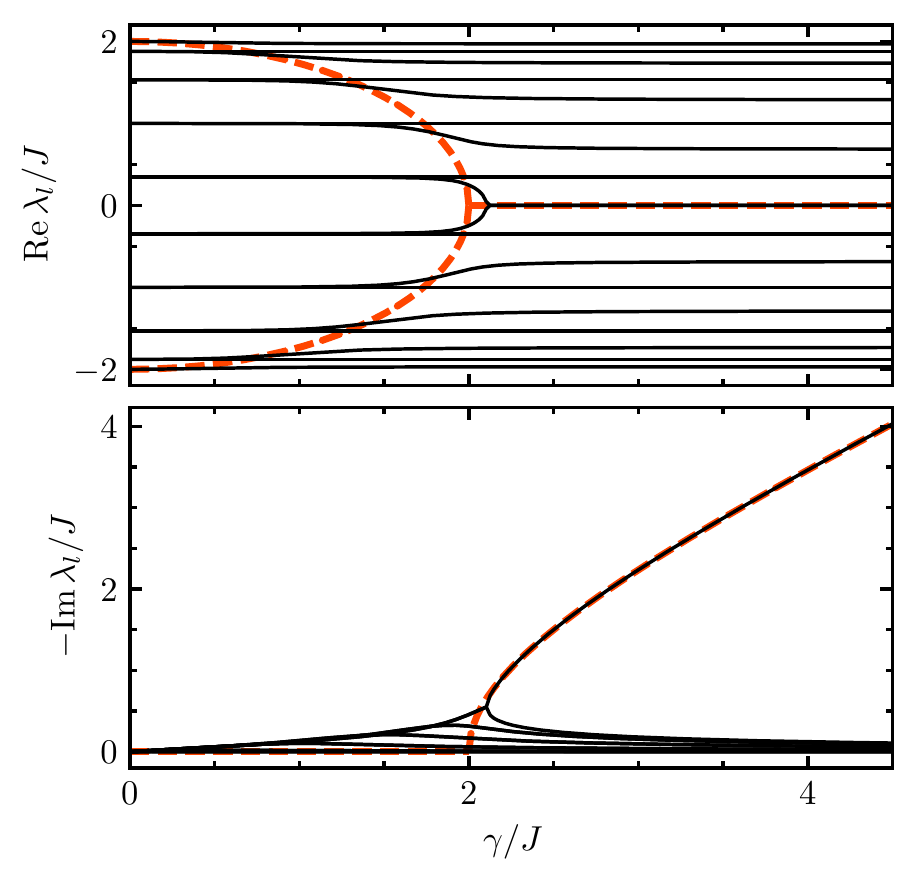}
\caption{%
Real (upper panel, black lines) and imaginary part (lower panel, black lines) of the eigenvalues $\lambda_l$ of the non-Hermitian effective Hamiltonian $\widetilde H$ as a function of $\gamma$, for $L = 18$ and periodic boundary conditions. The  red dashed lines indicate the analytical solutions~\eqref{eq:S_deloc_lambda}, respectively~\eqref{eq:S_locLength}.
}
\label{fig:S_spectrum} 
\end{figure}

We consider the retarded Green's function in frequency space $G_{i,j}(\omega)$ for a noninteracting system on a lattice of size $L$, described by the Hamiltonian~\eqref{eq:Hamiltonian-lattice} (with $U=0$) and in the presence of a localized loss $\gamma$.
The spectrum of $G_{i,j}(\omega)$  is related to the one of the non-Hermitian effective Hamiltonian $\widetilde H$ \eqref{eq:Heff} as a consequence of the following relationship: 
\begin{equation}
G^{-1}_{i,j}(\omega) = (\omega  + \ii \epsilon) \delta_{ij} - \widetilde H_{i j},
\end{equation}
with $\epsilon $ an infinitesimal positive quantity ensuring causality.
Since $\widetilde{H}$ is a non-Hermitian Hamiltonian, its eigenvalues $\lambda_l$ are complex. The real and imaginary parts of the spectrum for a system of finite size ($L=18$) are depicted in Fig.~\ref{fig:S_spectrum} as a function of $\gamma$.
The real part of the eigenvalues lies within $-2J$ and $ 2J$, i.e., in the same energy band as of the system in absence of the impurity. The values of the imaginary parts are typically of order $\sim \gamma /L$. 
Remarkably, the spectrum reorganizes at $\gamma_c = 2J$, and for $\gamma > \gamma_c$ a single eigenvalue acquires an imaginary part of order $\sim \gamma$, much larger than the other modes (see Fig.~\ref{fig:S_spectrum}, lower panel). The remaining eigenvalues possess a finite imaginary part, which becomes increasingly small upon increasing $\gamma$. 
Moreover, some eigenvalues have a vanishing imaginary part for any value of $\gamma$: these correspond to the dissipationless subspace discussed in Ref.~\onlinecite{Kepesidis2012} (see also discussion in Sec.~\ref{sec:timeregimes}).
For increasing system sizes $L$, the real and imaginary parts of the eigenvalue for the single distinct mode approach the red dashed lines in Fig.~\ref{fig:S_spectrum} (see discussion below). In contrast, the imaginary parts of all other modes decrease as $\sim 1/L$ upon increasing the system size. In fact, the sum of all the eigenvalues must satisfy $\sum_l \lambda_l = - \ii \gamma$ (cf. Eq.~\eqref{eq:Heff}), i.e., the imaginary parts of all eigenvalues must add up to $-\gamma$. Therefore, upon increasing $L$, the number of modes grows, while the contribution of each mode scales as $1/L$ in order to yield a finite sum (except for the outstanding mode, which contributes with a finite value to the sum).

In the real part of the spectrum (see Fig.~\ref{fig:S_spectrum}), a successive shift of frequencies takes place upon increasing $\gamma$, which ends at $\gamma = \gamma_c$. These shifts follow a semicircle pattern, corresponding to the energies of a specific set of delocalized solutions, as discussed further below. At $\gamma_c$, two previously nondegenerate values merge at zero frequency. This value corresponds to the mode with finite imaginary part.

\begin{figure}[t!]
\centering
\includegraphics[width = 0.45 \textwidth]{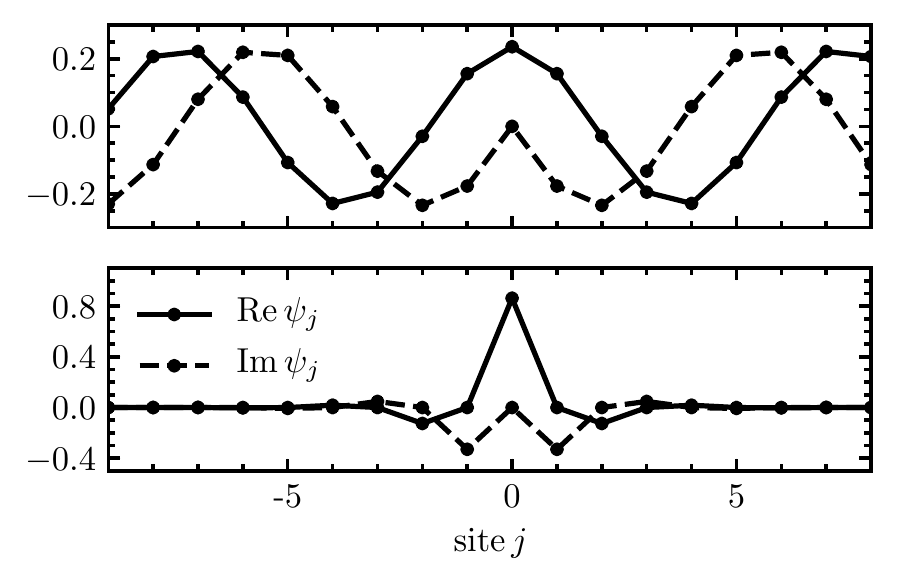}
\caption{%
Upper panel: Detail of the delocalized solution~\eqref{eq:S_homo_kappa} for $\gamma = 1.5 J$.
Lower panel: Detail of the localized-state solution~\eqref{eq:S_loc_kappa}, for $\gamma = 3 J$.
}
\label{fig:wf} 
\end{figure}

In order to further investigate the nature of this transition, we numerically inspected the corresponding eigenvectors.
The single mode acquiring a large imaginary eigenvalue turns out to be a localized state, exponentially localized at the loss site and only present for $\gamma > \gamma_c$ (see Fig.~\ref{fig:wf}, lower panel), while the other modes correspond to delocalized scattering states.

The spectral properties described so far can be interpreted as an incarnation of the QZE. In fact, the imaginary part (i.e., the inverse lifetime) of the delocalized modes features a nonmonotonic dependence on $\gamma$, as shown in Fig.~\ref{fig:S_spectrum}. On the converse, the localized mode is the fastest decaying mode and is related to the emptying of the loss site.

We further substantiate the nature of the localized state by solving analytically of the scattering problem defined by the non-Hermitian Schr{\"o}dinger equation
\begin{equation}
\label{eq:S_def_nonherm_scatt} 
\ii \del_t \psi_j = \sum_k \widetilde H_{j k} \psi_k
\end{equation}
for an infinitely large ($L=+\infty$) system. Exponentially localized solutions are found by using the ansatz
\begin{equation}
\psi_j = A \ee^{-\kappa |j| - \ii \lambda t}		,\qquad \kappa, \lambda \in \mathbb{C}	.
\label{eq:loct:analyticalAnsatz} 
\end{equation}
By writing $\kappa = \kappa_R + \ii \kappa_I$, we find the conditions 
\begin{equation}
\cosh{\kappa_R}\, \sin \kappa_I  = \frac{\gamma}{2 J } ,  \qquad
\sinh{\kappa_R}\, \cos \kappa_I  = 0,
\label{eq:loct:analyCond}
\end{equation}
while the eigenvalue $\lambda$ is given by
\begin{equation}
\lambda = - 2J \ee^{-\kappa} - \ii \gamma 	.
\label{eq:loct:eigenvalues} 
\end{equation}
Equations~\eqref{eq:loct:analyCond} admit two types of solutions. For $\gamma < \gamma_c = 2J$ two solutions exist, with $\kappa_R = 0$ and 
\begin{equation}
\kappa_I^{(1)} = \arcsin \frac{\gamma}{2 J}, \qquad \kappa_I^{(2)} = \pi - \kappa_I^{(1)}, 
\label{eq:S_homo_kappa} 
\end{equation}
which corresponds to the eigenvalues
\begin{equation}
\lambda^{(1)} = - \sqrt{\gamma_c^2 - \gamma^2},  \qquad \lambda^{(2)} = \sqrt{\gamma_c^2 - \gamma^2}, 
\label{eq:S_deloc_lambda} 
\end{equation}
indicated by the red dashed lines in Fig.~\ref{fig:S_spectrum}, for $\gamma < \gamma_c$.
These solutions are delocalized since $\kappa_R = 0$, and correspond to the ones experimentally investigated in Ref.~\onlinecite{Muellers2018}.  The solution $\psi^{(1)}$ is shown in Fig.~\ref{fig:wf} (upper panel). The shift of frequencies seen in the real part of the spectrum for $\gamma < \gamma _c$ follows the semicircle described by Eqs.~\eqref{eq:S_deloc_lambda}, see Fig.~\ref{fig:S_spectrum} (upper panel).

For $\gamma > \gamma_c$, instead, a single solution is found for which $\kappa_I = \pi/2 $ and
\begin{equation}
\kappa_R = \ln \left[  \frac{1}{2 J} \left( \gamma +  \sqrt{\gamma^2 - \gamma_c^2} \right) \right]	,
\label{eq:S_loc_kappa} 
\end{equation}
thus describing a localized function, shown in Fig.~\ref{fig:wf} (lower panel), with localization length $\kappa_R^{-1}$, decreasing with $\gamma$.  The corresponding eigenvalue $\lambda$ reads
\begin{equation}
\re \, \lambda = 0 , \qquad \im \, \lambda = -  \sqrt{\gamma^2 - \gamma_c^2} \label{eq:S_locLength} ,
\end{equation}
as indicated in Fig.~\ref{fig:S_spectrum}, for $\gamma > \gamma_c$, by the red dashed lines, and it shows perfect agreement with the finite-size spectrum, with  corrections being important at $\gamma\sim \gamma_c$. 
The two solutions of Eq.~\eqref{eq:S_homo_kappa} coalesce at $\gamma_c$, which therefore constitutes an exceptional point of $\widetilde H$~\cite{Heiss2012}.  
We remark that a localized state appears only in the lattice model while the continuum analog of Eq.~\eqref{eq:S_def_nonherm_scatt} does not support a localized solution.

The spectral properties discussed in this appendix connect to several previous works on systems subject to localized loss. In Ref.~\onlinecite{Muellers2018} similar homogeneous solutions were found and shown to be related to coherent perfect absorption, which was experimentally realized in a BEC. In Ref.~\onlinecite{Li2019}, this spectral transition for the two-site model ($L=2$ in Eq.~\eqref{eq:Heff}) has been experimentally observed in a Fermi gas.
In Ref.~\onlinecite{Jung1999} the spectral reorganization of systems coupled locally to external modes has been discussed for a general case, in particular pointing out the emergence of exceptional points and the relation of these features to phase transitions. A similar phase transition has been also described in Ref.~\onlinecite{Demler2018} for a one-dimensional BEC with a localized loss. 
In passing, we notice that Ref.~\onlinecite{Haque2019} appeared recently, pointing out the same transition described here.

\section{Currents}
\label{sec:S_currents}
Starting from the definition
\begin{equation}
j(x) =\frac{1}{m} \im \partial_y C(x,y)  |_{y=x},
\end{equation}
we obtain, using Eq.~\eqref{eq:correlationsfree},
\begin{align}
\label{eq:current-derivation}
\im \partial_y C(x,y)  |_{y=x} = \sgn(x)  \int_k  k \left[ 
\abs{r_k}^2 + \re( r_k )
  \right] n_{0,k}.
\end{align}
From Eq.~\eqref{eq:romega} we obtain
\begin{equation}
|r_k|^2 + \re( r_k) = - \frac{\eta_k}{2}		,
\end{equation}
which, together with Eq.~\eqref{eq:current-derivation}, and by comparison with Eq.~\eqref{eq:nimp}, yields Eq.~\eqref{eq:current} in the main text. Note that, in the case of a potential barrier (i.e., by replacing $\gamma \rightarrow \ii g  $), the expression $|r_k|^2 + \re( r_k)$ vanishes, consistently with the fact that no currents are expected for a coherent impurity. 

\section{Momentum distribution}
\label{sec:S_momentumDistribution} 

In this appendix we sketch the derivation of the momentum distribution formulas of Secs.~\ref{sec:nK}.
The momentum distribution can be obtained from the Fourier transform of Eqs.~\eqref{eq:correlation-retarded} and~\eqref{eq:Gxx} as:
\begin{multline}
\label{eq:momentum-distro}
\langle \psi^\dagger(k,t) \psi(k,t) \rangle = \int_q n_{0,q} \Big\{ |G_0(k,t-t_0)|^2  \delta^2_{k,-q}  \\
 + \delta_{k,-q} 2\text{Re}\left[G_0^*(k,t-t_0) \widetilde G(k,-q,t-t_0)\right] \\ + |\widetilde G(k,-q,t-t_0)|^2  \Big\}, 
\end{multline}
where 
\begin{equation}
 \widetilde G(k,k',\omega) =  \widetilde r(\omega) G_0(k ,\omega) G_0(k' ,\omega),
\end{equation}
with $\widetilde r(\omega) = i |v(\omega)| r(\omega)$ and $r(\omega)$ given in Eq.~\eqref{eq:romega},  $ v(\omega)$ the group velocity, and $G_0(k,\omega)$ the Fourier transform of Eq.~\eqref{eq:G0xx}.
A straightforward calculation yields
\begin{multline}
\langle   \psi^\dagger(k,t)   \psi(k,t) \rangle = 2 \pi n_{0,k} \\
\times  \left[ 
\delta(k=0) + \delta(\omega = 0) \abs{v(\epsilon_k)} 2 (r(\epsilon_k) + r^2(\epsilon_k) )
\right] ,
\end{multline} 
where the delta functions are evaluated at zero and account for a factor diverging as the system size in the limiting procedure of Eq.~\eqref{eq:momentum_distro_def} discussed in Sec.~\ref{sec:nK}.
Here, we can identify the loss probability $\eta_k = - 2 (r(\epsilon_k) + r^2(\epsilon_k))$.

\section{Bosonization: strong coupling limit}
\label{sec:suppl:strongcoupling} 

We derive the explicit form of the dissipative tunneling between two semi-infinite wires which are separated by a strong loss barrier. To this end, we model this configuration as a three-site system with localized loss at the central site: by integrating out the central site, we obtain an effective dissipative coupling between the remaining two sites.

We consider three sites labeled as $j=A,X,B$ and  described by the Hamiltonian $H=-J(\psi_X^\dagger \psi_{A} + \psi_B^\dagger \psi_X + h.c.) $ and Lindblad operators $\LB_j = \psi_j \delta_{j,X}$, implementing a localized loss at the central site $X$. Here, $\psi_{j}$, $\psi_{j}^\dagger$ denote fermionic operators. 
The total action is then given by $S[\psi_A,\psi_X,\psi_B]= S_0[\psi_A,\psi_B]+S_0[\psi_X]+S_\text{int}[\psi_A,\psi_X,\psi_B]$, with $S_0$ the actions for the decoupled sites and $S_\text{int}$ the action describing the tunneling produced by $H$.
The effective action $S_\text{eff}[\psi_A, \psi_B]$ coupling the sites $A$ and $B$ is obtained by integrating out the central site $X$ as
\begin{equation}
\int \mathcal{D}\psi_X\, \ee^{\ii S_0[\psi_X]+i S_\text{int}[\psi_{A}, \psi_X, \psi_B]} \sim \ee^{\ii  S_\text{eff}[\psi_A, \psi_B]},
\end{equation}
and yields
\begin{equation}
 \!\!S_\text{eff}=\! \int_\omega \left[ \ii  \frac{2 J^2 \gamma}{\omega^2 + \gamma^2} \bar \TB_q \TB_q - \left( \frac{J^2}{\omega + \ii \gamma} \bar \TB_q \TB_c +\text{c.c.}\right) \right],
\end{equation}
with $\TB_{c/q} = \psi_{c/q,A} + \psi_{c/q,B}$. To leading order in $\gamma^{-1}$ we obtain 
\begin{equation}
S_{D} = - \ii D \int_\omega \left[- 2 \TB_q^* \TB_q + \left( \TB_c^*\TB_q - \TB_q^* \TB_c \right) \right], \label{eq:supp:S_D} 
\end{equation}
with $D =  \gamma^{-1} J^2 $. The action $S_D$ corresponds exactly to the one obtained from a Lindblad operator of the form $\LB = \psi_{A} + \psi_B$~\cite{Sieberer_review}.
We therefore infer that the effect of a large but finite localized loss can be modeled as two disconnected leads coupled by a dissipative tunneling implemented by the Lindblad operator $\LB = \psi_A(x=0) + \psi_B(x=0)$ with strength $D= J^2/\gamma$,  with $\psi_{A,B}(x=0)$ corresponding to the fermionic annihilation operators at the tip points of the two semi-infinite parts $A$ and $B$.

\section{Derivation of RG equations in the Luttinger liquid description}
\label{app:derivation-RG-Luttinger} 

In this appendix we sketch the derivation of the RG equations~\eqref{eq:RG-bosonization}. In particular, we show how to compute the contributions to the effective action in Eq.~\eqref{eq:cumulants}.\\
The first cumulant $\langle S_\text{loss} \rangle_f = \langle S_\text{n} \rangle_f + \langle S_\text{b} \rangle_f$, with $S_\text{n}$ and $S_\text{b}$ defined in Eqs.~\eqref{eq:Keldysh_impurity}, reads
\begin{subequations}
\begin{align}
\left\langle   S_\text{n}[\theta^{s+f},\phi^{s+f}]   \right\rangle_f &=  S_\text{n}[\theta^{s},\phi^{s}] 	,\label{eq:noise-uncorrelated}\\
\left\langle   S_\text{b}[\theta^{s+f},\phi^{s+f}]   \right\rangle_f  & =  S_\text{b} [\theta^{s},\phi^{s}] \ee^{- \left\langle \theta_{c}^2(x,t) \right\rangle_f },
\end{align}
\end{subequations}
where ``f'' and ``s'' denote fast and slow fields, respectively. We notice that the noise action $S_\text{n}$ does not bear any renormalization as it only contains quantum fields. This first-order correction to $S_\text{b}$ can thus be calculated from~\eqref{eq:local_action_RG}, and it reads
\begin{equation}
\langle \theta_c^2(x,t)\rangle_f  = \int_{\Lambda\ee^{-\ell}<|\omega|<\Lambda}  \langle |\Theta_c|^2\rangle=  \ell\left(\frac{\kappa_0}{\kappa^2}+\frac{2\bar{T}}{\kappa}\right)	.
\end{equation}
The second cumulant in Eq.~\eqref{eq:cumulants} leads to a contribution $\delta S^{(2)}$ given by
\begin{multline}
    \delta S^{(2)} = \frac{1}{2} \bigg( \langle S_\text{n}^2 \rangle_f - \langle S_\text{n}\rangle_f^2 + \langle S_\text{b}^2 \rangle_f - \langle S_\text{b}\rangle_f^2 \\
    + 2\langle S_\text{n}S_\text{b}\rangle_f - 2\langle S_\text{n}\rangle_f \langle S_\text{b}\rangle_f \bigg),
\end{multline}
where, by using Eq.~\eqref{eq:noise-uncorrelated}, we immediately find that 
\begin{equation}
 \langle S_\text{n}^2 \rangle_f - \langle S_\text{n} \rangle_f^2  = 0, 
\end{equation}
since the noise term only depends on quantum fields. Among the other terms in $\delta S^{(2)}$, we recognize two classes of terms: (i) terms containing higher harmonics nonlinear terms, such as $\ee^{i \sqrt{2} \phi_q}, \cos(\sqrt{2}\theta_c), \dots$, and (ii) terms renormalizing $S^\Theta_0$. In particular, no further corrections to $S_\text{n}$ and $S_\text{b}$ are generated at second order. In the following, we will sort out the contributions to $S^\Theta_0$.
The cross terms $\sim S_\text{n} S_\text{b}$
also do not generate any contribution to $S_0$.
Thus, the only renormalization at second order of $S_0$ originates from the backscattering action.
We will resort to the short notation $\theta(x=0, t) \equiv \Theta(t) \equiv \Theta$ and $\theta(x=0, t') \equiv \Theta'$ for the sake of readability. In the following, $\alpha=\sqrt{2}$. 
The squared backscattering  action reads:
\begin{widetext}
\begin{align}
 &S_\text{b}^2  = -4\gamma^2 \int_{t,t'} \Bigg\{  \nonumber \\
& \frac{1}{8}  \Big[ \cos{\alpha(\Theta_c+\Theta_c'+\Theta_q+\Theta_q')}+\cos{\alpha(\Theta_c-\Theta_c'+\Theta_q+\Theta_q')}  + \cos{\alpha(\Theta_c+\Theta_c'-\Theta_q+\Theta_q')}+\cos{\alpha(\Theta_c+\Theta_c'+\Theta_q-\Theta_q')} \nonumber \\
    & + \cos{\alpha(\Theta_c-\Theta_c'-\Theta_q+\Theta_q')}+\cos{\alpha(\Theta_c-\Theta_c'+\Theta_q-\Theta_q')}  + \cos{\alpha(\Theta_c+\Theta_c'-\Theta_q-\Theta_q')}+\cos{\alpha(\Theta_c-\Theta_c'-\Theta_q-\Theta_q')}\Big]  \label{eq:quarticcosines}\\
    & + \frac{1}{2} \ee^{i \alpha (\phi_q + \phi_q')} \left[ \cos{\alpha(\Theta_c+\Theta_c')} + \cos{\alpha(\Theta_c-\Theta_c')} \right]  \label{eq:higherharmonics}\\
    &- \frac{1}{4} \ee^{i \alpha \phi_q} \Big[ \cos{\alpha(\Theta_c + \Theta_c' + \Theta_q')} + \cos{\alpha(\Theta_c - \Theta_c' + \Theta_q')} + \cos{\alpha(\Theta_c + \Theta_c' - \Theta_q')} + \cos{\alpha(\Theta_c - \Theta_c' - \Theta_q')} \Big] \nonumber \\
&    - \frac{1}{4} \ee^{i \alpha \phi_q'} \Big[ \cos{\alpha(\Theta_c' + \Theta_c + \Theta_q)} + \cos{\alpha(\Theta_c' - \Theta_c + \Theta_q)}  + \cos{\alpha(\Theta_c' + \Theta_c - \Theta_q)} + \cos{\alpha(\Theta_c' - \Theta_c - \Theta_q)} \Big] \Bigg\}. \label{eq:eliminateclassical}
\end{align}
\end{widetext}
The term~\eqref{eq:higherharmonics} only contributes to higher nonlinear harmonics. The terms~\eqref{eq:quarticcosines} generate terms $\propto \Theta_q^2$ and  $\propto \Theta_c \Theta_q$, while \eqref{eq:eliminateclassical} is needed to eliminate terms $\propto \Theta_c^2$ in order to preserve the Keldysh causality structure.
After decomposing the field in slow and fast components, the individual terms in~\eqref{eq:quarticcosines} read
\begin{multline}
    \langle \cos{\alpha(\Theta_c\pm\Theta_c'\pm\Theta_q\pm\Theta_q')}\rangle_f =\\
 \cos{\alpha(\Theta_c^s\pm\Theta_c^{'s}\pm\Theta_q^s\pm\Theta_q^{'s})} \ee^{-\frac{\alpha^2}{2}\langle (\Theta_c^f\pm\Theta_c^{'f}\pm\Theta_q^f\pm\Theta_q^{'f})^2\rangle_f}.
\end{multline}
The cosines containing slow fields can, in turn, be expanded around their mean,
\begin{multline}
    \cos{\alpha(\Theta_c^s\pm\Theta_c^{s'}\pm\Theta_q^s\pm\Theta_q^{s'})} =\\
    :\!\!\cos{\alpha(\Theta_c^s\pm\Theta_c^{s'}\pm\Theta_q^s\pm\Theta_q^{s'})}\!\!: \ee^{-\frac{\alpha^2}{2}\langle (\Theta_c^s\pm\Theta_c^{'s}\pm\Theta_q^f\pm\Theta_q^{'s})^2\rangle_{s}},
\end{multline}
where $:\dots:$ denotes normal ordering~\cite{DallaTorre2012}. The normal ordered cosines can then be safely expanded to second order as $:\!\!\cos{\Theta}\!\!:  = 1 - \Theta^2/2 + \mathcal{O}((\Theta-\langle \Theta\rangle)^4)$. Recombining the fast and slow correlators, the expansion of each individual term in~\eqref{eq:quarticcosines} takes the form
\begin{multline}
  \langle \cos{\alpha(\Theta_c\pm\Theta_c'\pm\Theta_q\pm\Theta_q')}\rangle_{f} =\\ 
 \!\left[\!1\!-\!\frac{\alpha^2}{2}(\Theta_c^s \!\pm\! \Theta_c^{'s} \!\pm\! \Theta_q^s \!\pm\! \Theta_q^{'s})^2\right]\! \ee^{-\frac{\alpha^2}{2}\langle (\Theta_c\pm\Theta_c^{'}\pm\Theta_q\pm\Theta_q^{'})^2\rangle_{f\!+\!s}}.
\end{multline}
As the integrand of $S_\text{b}^2$ is symmetric under exchanging $t $ with $t'$, the integration can be restricted to $t>t', \int_{-\infty}^\infty \dd{t} \int_{-\infty}^\infty \dd{t'} = 2 \int_{-\infty}^\infty \dd{t} \int_{-\infty}^t \dd{t'}$. Furthermore, it is useful to switch to central and relative time coordinates, $ t_c  = t + t'$  and $  t_r  = (t-t')/2$, respectively, which allows for an expansion of the fields as
\begin{subequations}
\begin{align}
    \Theta_{c/q}(t)+\Theta_{c/q}(t') & = 2 \Theta_{c/q} (t_c) + \mathcal{O}(t_r^2) ,\\
    \Theta_{c/q}(t)-\Theta_{c/q}(t') & = t_r \frac{\dd \Theta_{c/q}}{\dd t_c} (t_c) + \mathcal{O}(t_r^2),
\end{align}
\end{subequations}
assuming quickly decaying field correlators for large $t_r$. In the following, we define $C_K(t) = \langle \Theta_c(t)\Theta_c(0)\rangle$ and  $C_R(t) = \langle \Theta_c(t) \Theta_q(0)\rangle$, and by $C^f_{K,R}(t)$ the corresponding correlations averaged only on the fast fields. When plugged back into $\delta S^{(2)}$, each term in Eq.~\eqref{eq:eliminateclassical} has to be paired with a corresponding term originating from $\langle S_\text{b}\rangle_{f}^2 = S_\text{b}^2[\Theta^s] \exp[{-2\alpha^2 C_K^f(0)}]$.  All additive constants in the action will be dropped, as inconsequential for the dynamics. Finally, one obtains:
\begin{widetext}
\begin{align}
    \delta S^{(2)} = \frac{\gamma^2 \alpha^2}{4} \int_{t_c, t_r} & \Bigg\{ (2 \Theta_{c}^s + 2 \Theta_{q}^s)^2 \ee^{-\alpha^2 (C_K(0)+C_K(t_r)+C_R(t_r))} \left(1 - \ee^{-\alpha^2 (C_K^f(0)-C_K^f(t_r)-C_R^f(t_r))} \right) \nonumber \\
    & + (t_r \partial_{t_c} \Theta_{c}^s + 2 \Theta_{q}^s)^2 \ee^{-\alpha^2 (C_K(0)-C_K(t_r)+C_R(t_r))} \left(1 - \ee^{-\alpha^2 (C_K^f(0)+C_K^f(t_r)-C_R^f(t_r))} \right) \nonumber \\ 
    & + (2 \Theta_{c}^s - t_r \partial_{t_c} \Theta_{q}^s)^2 \ee^{-\alpha^2 (C_K(0)+C_K(t_r)+C_R(t_r))} \left(1 - \ee^{-\alpha^2 (C_K^f(0)-C_K^f(t_r)-C_R^f(t_r))} \right)\nonumber  \\
    & + (2 \Theta_{c}^s + t_r \partial_{t_c} \Theta_{q}^s)^2 \ee^{-\alpha^2 (C_K(0)+C_K(t_r)-C_R(t_r))} \left(1 - \ee^{-\alpha^2 (C_K^f(0)-C_K^f(t_r)+C_R^f(t_r))} \right)\nonumber  \\
    & + (t_r \partial_{t_c} \Theta_{c}^s - t_r \partial_{t_c} \theta_{q}^s)^2 \ee^{-\alpha^2 (C_K(0)-C_K(t_r)+C_R(t_r))} \left(1 - \ee^{-\alpha^2 (C_K^f(0)+C_K^f(t_r)-C_R^f(t_r))} \right)\nonumber  \\
    & + (t_r \partial_{t_c} \Theta_{c}^s + t_r \partial_{t_c} \Theta_{q}^s)^2 \ee^{-\alpha^2 (C_K(0)-C_K(t_r)-C_R(t_r))} \left(1 - \ee^{-\alpha^2 (C_K^f(0)+C_K^f(t_r)+C_R^f(t_r))} \right) \nonumber \\
    & + (2 \Theta_{c}^s - 2 \Theta_{q}^s)^2 \ee^{-\alpha^2 (C_K(0)+C_K(t_r)-C_R(t_r))} \left(1 - \ee^{-\alpha^2 (C_K^f(0)-C_K^f(t_r)+C_R^f(t_r))} \right)\nonumber  \\
    & + (t_r \partial_{t_c} \theta_{c}^s - 2 \Theta_{q}^s)^2 \ee^{-\alpha^2 (C_K(0)-C_K(t_r)-C_R(t_r))} \left(1 - \ee^{-\alpha^2 (C_K^f(0)+C_K^f(t_r)+C_R^f(t_r))} \right) \Bigg\} \nonumber \\
    & + \text{less relevant terms}.
\end{align}
\end{widetext}
The multiplicative factors $1-\ee^{-C_K^f(0)+\dots}$ are already of order $\ell = \ln{\Lambda/\Lambda'}$, which means that, upon rescaling $\Lambda' \rightarrow \Lambda$, all slow fields in this expression can be replaced by the full fields. To further simplify the expression, we can drop all terms proportional to $\ee^{-\alpha^2 (C_K(0)+C_K(t_r))}$, as they decay much faster compared to $\ee^{-\alpha^2 (C_K(0)-C_K(t_r))}$. The remaining terms can be reorganized as
\begin{equation}
\label{eq:action-contribution}
\delta S^{(2)}  = 2\gamma^2 \alpha^2 \ell \int_{t>0}  \left( J_T \Theta_q^2 +  J_\kappa  \Theta_{q} i\partial_{t} \Theta_{c} \right),
\end{equation}
where the integrals $J_T$ and $J_\kappa$ are defined as
\begin{multline}
    J_T  = \text{Re} \int_{t>0}   \ee^{-\alpha^2(C_K(0)-C_K(r)-C_R(t))} \times \\ \times \frac{1}{\ell}\left( 1 - \ee^{-\alpha^2(C^f_K(0)+C^f_K(t)+C^f_R(t))} \right),
\end{multline}
and
\begin{multline}
    J_\kappa  = - \text{Im} \int_{t>0} t  \ee^{-\alpha^2(C_K(0)-C_K(r)-C_R(t))} \times \\ \times \frac{1}{\ell}\left( 1 - \ee^{-\alpha^2(C^f_K(0)+C^f_K(t)+C^f_R(t))} \right).
\end{multline} 
\begin{figure}[t!]
\centering
\includegraphics[width = 0.4 \textwidth]{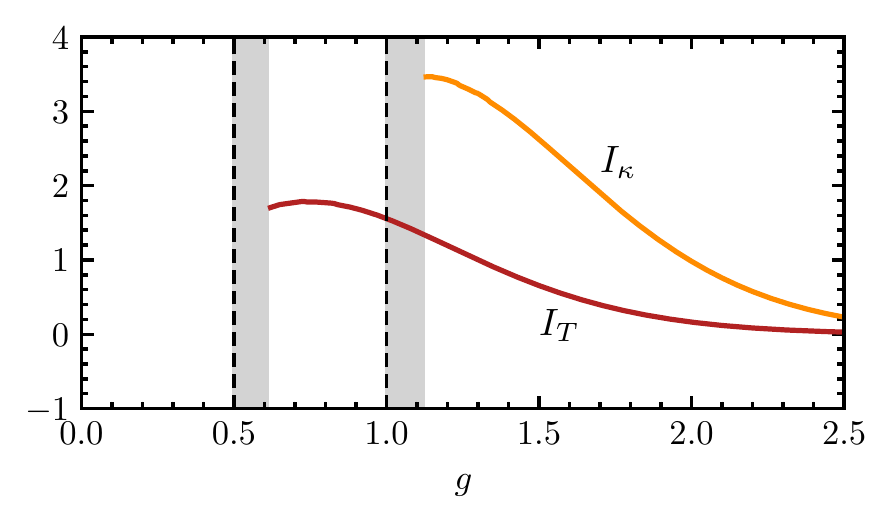}
\caption{%
(Color online). Integrals $I_T$ and $I_\kappa$ as a function of the Luttinger parameter $g$ for $T=0$. The gray-shaded areas indicate insufficient numerical convergence.}
\label{fig:rg_integrals} 
\end{figure}
Finally, by comparing Eq.~\eqref{eq:action-contribution} with the quadratic action~\eqref{eq:local_action_RG}, one obtains the RG equations~\eqref{eq:RG-bosonization} in the main text, where the explicit form of the integrals $I_{T,\kappa}$ reads
\begin{subequations}
\begin{align}
\label{eq:IT}
I_T &=  \pi \text{Re}
\bigg[
\int_0^\infty \dd t\,  \ee^{f(t)} \,
h(t)
\bigg], \\
\label{eq:Ikappa}
I_\kappa &= - 2\pi \text{Im}
\bigg[
\int_0^\infty \dd t\,  t\,  \ee^{f(t)} \,  h(t)
\bigg],
\end{align}
\end{subequations}
with the functions $h(t)$ and $f(t)$ given by
\begin{equation}
h(t) =  \left[ 2\left(\frac{\kappa_0}{\kappa^2} +\frac{2\bar{T}}{\kappa}\right)(1+\cos t) - \frac{i}{\kappa}\sin t\right]	,
\end{equation}
and
\begin{multline}
f(t) = \frac{2\kappa_0}{\kappa^2} \left[-\gamma_{EM} +\text{Ci}(t) -\log t \right] \\ +\frac{4\bar{T}}{\kappa}\left[1-\cos t - t\,\text{Si}(t)\right] -\frac{2i}{\kappa}\text{Si}(t),
\end{multline}
with $\gamma_{EM}$ the Euler-Mascheroni constant, and $\text{Ci}(t), \text{Si}(t)$ the cosine integral and sine integral functions, respectively.
The values of the integrals $I_{T,\kappa}$ for $T=0$ and as a function of the Luttinger parameter $g$ are shown in Fig.~\ref{fig:rg_integrals}. $I_T$ converges down to $g=1/2$, while $I_\kappa$ converges down to $g =1$, entailing that the RG flow with $T=0$ as initial condition  is not applicable for $g<1$. The gray-shaded areas correspond to regions where the numerical estimate of the integrals becomes unstable.
%

\bibliography{biblio}

\end{document}